\documentclass[12pt]{article}
\pdfoutput=1
\usepackage{amsmath}
\usepackage{amssymb}
\usepackage{epsfig}
\usepackage{cite}
\usepackage{fontenc}
\usepackage{graphicx}
\usepackage{float}
\usepackage[lofdepth,lotdepth,caption=false]{subfig}
\usepackage{color}
\usepackage{booktabs}
\usepackage{tabularx}
\usepackage{multirow}
\usepackage{longtable}
\usepackage{lscape}
\usepackage{dcolumn}
\usepackage{rotating}
\usepackage[normalem]{ulem}
\definecolor{darkgreen}{cmyk}{1,0,1,0.4}
\definecolor{brown}{cmyk}{0,0.8,1,0.2}
\definecolor{darkred}{cmyk}{0,1,1,0.2}

\allowdisplaybreaks[4] 
\makeatletter
\renewcommand{\fnum@table}{\textbf{\tablename~\thetable}}
\renewcommand{\fnum@figure}{\textbf{\figurename~\thefigure}}
\makeatother

\newcounter{myenumi}

\renewcommand{\themyenumi}{\roman{myenumi}}
{\end{list}}

\newlength{\myem}
\newcounter{mysubequation}[equation]

\makeatletter
\renewcommand{\section}{\@startsection{section}{1}{0em}{-\baselineskip}%
{\baselineskip}{\normalfont\large\bfseries}}
\renewcommand{\subsection}%
{\@startsection{subsection}{2}{0em}{-0.7\baselineskip}%
{0.7\baselineskip}{\normalfont\bfseries}}
\makeatother

\newcommand{\bi}{\begin{itemize}}
\newcommand{\ei}{\end{itemize}}

\def\beq{\begin{equation}}
\def\eeq{\end{equation}}

\newcommand{\bea}{\begin{eqnarray}}
\newcommand{\eea}{\end{eqnarray}}

\newcommand{\pme}{P_{\mu e}}

\newcommand{\ie}{{\it i.e.}}

\newcommand{\eet}{\varepsilon_{e\tau}}
\newcommand{\emt}{\varepsilon_{\mu\tau}}
\newcommand{\ett}{\varepsilon_{\tau\tau}}
\newcommand{\eee}{\varepsilon_{ee}}
\newcommand{\eem}{\varepsilon_{e\mu}}
\newcommand{\emm}{\varepsilon_{\mu\mu}}

\newcommand{\eetp}{\varphi_{e\tau}}
\newcommand{\emtp}{\varphi_{\mu\tau}}
\newcommand{\eemp}{\varphi_{e\mu}}
\def\epsilon{\varepsilon}

\newcommand{\chisq}{\ensuremath{\chi^2}}

\newcommand{\ttok}{{\sc T2K}}
\newcommand{\dune}{{\sc DUNE}}

\newcommand{\nova }{{\sc NOvA}}

\newcommand{\ttohk}{{\sc T2HK}}

\newcommand\sch{Schr$\ddot{\rm o}$dinger~}

\def\<{\langle}
\def\>{\rangle}

\def\lsim{\mathrel{\rlap{\lower4pt\hbox{\hskip1pt$\sim$}}
    \raise1pt\hbox{$<$}}}         
\def\gsim{\mathrel{\rlap{\lower4pt\hbox{\hskip1pt$\sim$}}
    \raise1pt\hbox{$>$}}}         

\begin{document}
%

\begin{titlepage}

\renewcommand{\thefootnote}{\alph{footnote}}

\vspace*{-3.cm}
\begin{flushright}

\end{flushright}


\renewcommand{\thefootnote}{\fnsymbol{footnote}}
\setcounter{footnote}{-1}

{
\begin{center}
{\large \bf
{{Non-standard interactions spoiling the CP violation sensitivity at DUNE and other long baseline experiments}
}
\\[0.2cm]
}
\end{center}
}

\renewcommand{\thefootnote}{\alph{footnote}}

\vspace*{.8cm}
\vspace*{.3cm}
{\begin{center} 
            {{\sf 
                Mehedi Masud$^\star$~\footnote[1]{\makebox[1.cm]{Email:}
                masud@hri.res.in}, and 
                Poonam Mehta$^{\ddagger}$~\footnote[2]{\makebox[1.cm]{Email:}
                pm@jnu.ac.in}
               }}
\end{center}}
\vspace*{0cm}
{\it 
\begin{center}
\footnotemark[1]%
$^\star$\, Harish-Chandra Research Institute, Chhatnag Road, Jhunsi, Allahabad 211 019, India
\footnotemark[2]%
$^\ddagger$ \, School of Physical Sciences, Jawaharlal Nehru University, 
      New Delhi 110067, India
\end{center}}

\vspace*{1.5cm}

\begin{center}
{\Large \today}
\end{center}

{\Large 
\bf
\begin{center} Abstract 
\end{center} 
 }
 It is by now established that neutrino oscillations occur due to non-zero masses and parameters in the leptonic mixing matrix.  The extraction of oscillation parameters may be complicated due to subleading effects such as non-standard neutrino interactions (NSI) and one needs to have a fresh look how a particular parameter value is inferred from experimental data. In the present work, we focus on an important  parameter entering the oscillation framework - the leptonic CP violating phase $\delta$, about which we know very little. We demonstrate that the sensitivity to CP violation gets significantly impacted due to NSI effects for the upcoming long baseline experiment, Deep Underground Neutrino Experiment (\dune). 
We also draw a comparison with the sensitivities of other ongoing neutrino beam experiments such 
 as \nova\, and \ttok\, as well as a future generation  experiment, \ttohk.

 \vspace*{.5cm}

\end{titlepage}

\newpage

\renewcommand{\thefootnote}{\arabic{footnote}}
\setcounter{footnote}{0}


\section{Introduction}
\label{sec:intro}

The discovery of neutrino oscillations implies that neutrinos have masses and  mix among three active flavours. If neutrinos have masses 
then  the leptonic charged current interactions exhibit mixing and 
CP~\footnote{CP stands for charge conjugation and parity discrete symmetry operation} violation in much the same way as in the quark sector
 ~\cite{Wolfenstein:1964ks,Branco:2011zb}. 
 Within the Standard  Model (SM),  CP violation arises naturally via the 
  Dirac phase, $\delta$ in the three flavour case as concocted by Kobayashi and Maskawa~\cite{km}. 
  It was suggested~\cite{Barger:1980jm,Nunokawa:2007qh}, that a measurement of $\delta$ was possible through neutrino oscillations.  The value of $\delta$ could very well be zero, maximal ($\delta \sim \pm \pi/2$) or 
non-maximal ($\delta \neq \pm \pi/2$)~\cite{Farzan:2006vj}. The extraction of the value of CP phase is plagued with the matter-induced fake CP 
violating  effects which makes its measurement very challenging even in the case of SM~\cite{Arafune:1997hd,Ohlsson:2013ip}.  Additionally, we need to know the ordering of the neutrino masses and also  
the octant of $\theta_{23}$  in order to have a clear understanding of the mixing phenomena.

 Deep Underground Neutrino  Experiment (\dune) is one of the most promising upcoming
  long baseline experiments that is planned to offer maximal sensitivity to uncover 
the value of $\delta$~\cite{Marciano:2006uc,Bass:2013vcg,Acciarri:2015uup,Acciarri:2016ooe,Acciarri:2016crz}. The baseline of 1300 km is expected to deliver optimal sensitivity to CP violation and is well-suited to address the question of neutrino mass hierarchy~\cite{Qian:2015waa}. 
 Sensitivity studies leading to optimal configurations have been carried out for \dune\, stand alone as well as in conjunction with other long baseline experiments and atmospheric neutrinos and it was shown that the CP violation can be established for $\sim 80\%$ of the CP phase values~\cite{Barger:2013rha,Barger:2014dfa} 
 (see also \cite{Ghosh:2014rna}) 
 under favourable conditions ($\delta $ around $ \pm \pi/2$). Of course, all these studies assume the SM where the only source of CP violation is the Dirac CP phase. 
Other interesting proposals to look for leptonic CP violation include the proposed Tokai to HyperKamiokande (\ttohk) experiment~\cite{Abe:2015zbg}  and MuOn-decay MEdium baseline NeuTrino beam (MOMENT) experiment which exploits neutrinos from muon decay~\cite{Blennow:2015cmn}.  

Since there are clear hints of  existence of new physics beyond the SM,
it is likely that the SM does not provide a complete 
description of CP violation in nature. A particular new physics scenario would introduce additional sources of CP violating effects 
  in addition to the lone CP violating source ($\delta$) of the SM (see~\cite{Farzan:2015doa,Farzan:2015hkd} for models giving rise to NSI) and  this could change the relationship of measured quantities to the
   CP violating parameter of the SM~\cite{GonzalezGarcia:2001mp}. In presence of new physics, we need to reassess the conclusions drawn in connection to CP violation. This was discussed in the context of Neutrino Factory 
in~\cite{Winter:2008eg} (see also~\cite{Chatterjee:2014gxa} in context of atmospheric neutrinos) which led to the conclusion that a new physics effect might be mis-interpreted as the 
canonical Dirac CP violation or vice-versa. 
Concerning the value of the CP phase $\delta$, there was a mild hint of its value from
 the global analyses of neutrino data~\cite{Forero:2014bxa} and in recent years, the
 \ttok~\cite{Abe:2015awa}, \nova~\cite{Adamson:2016tbq}\, as well as 
 SK atmospheric events~\cite{sktalk} indicate nearly maximal value 
 ($\delta \sim -\pi/2$) for the CP phase  although presently the 
 significance of this result is below $3\sigma$. If the significance improves in future,
  it is very important to rule out alternative mechanisms such as  NSI~\cite{Forero:2016cmb} (also see Ref.~\cite{Ohlsson:2012kf,Miranda:2015dra} for reviews and references therein) or CPT violation~\cite{Datta2004356,Chatterjee:2014oda}
  or additional 
  sterile neutrinos (see~\cite{Gandhi:2015xza,Berryman:2015nua}) that could very well be consistent with data.

In a recent work~\cite{Masud:2015xva} (see also~\cite{Friedland:2012tq,Coloma:2015kiu,deGouvea:2015ndi,Forero:2016cmb,Liao:2016hsa,Huitu:2016bmb,Bakhti:2016prn}), we studied the impact of flavour diagonal and flavour changing neutral current (NC) non-standard interactions (NSI) during propagation on CP measurements at long baselines using \dune\, as an example.
We discussed the role of individual and collective NSI parameters on the CP measurements. We restricted ourselves to a discussion of CP asymmetries using solely the probability for the electron appearance channel $\pme$ and showed how NSI effects translate at the level of event rates. Since then there has been a lot of activity~\cite{Coloma:2015kiu,deGouvea:2015ndi,Forero:2016cmb,Liao:2016hsa,Huitu:2016bmb,Bakhti:2016prn} in this direction exploring effects due to propagation NSI at long baselines.  
 In~\cite{Coloma:2015kiu}, the goal was to discuss possible improvement of bounds on the  NSI  parameters using different channels at long baseline experiments. 
  In~\cite{deGouvea:2015ndi} the correlation between different SI and NSI parameters and ways to distinguish scenarios of new physics (sterile neutrinos versus NSI) compared to the standard case was discussed. 
 In~\cite{Liao:2016hsa}, it was shown that  new degeneracies in presence of NSI at long baselines  can arise which complicate the determination of mass hierarchy, CP violation and octant of $\theta_{23}$. Ref.~\cite{Huitu:2016bmb} deals with constraining NSI parameters using long baseline experiments.
 Very recently, it has been suggested that the proposed MOMENT experiment due to its much smaller baseline  may prove helpful in solving the degeneracies due to NSI leading to unambiguous determination of CP violation~\cite{Bakhti:2016prn}.

While sensitivity studies have been carried out in presence of NSI in the context of \dune, we would like to stress that none of them deal with the precise impact of NSI on the sensitivity to CP violation at long baseline experiments. We 
 consider NSI terms whose strengths lie in the presently allowed limits (along with the phases associated with these terms which are presently unconstrained) and study the impact of individual and collective NSI terms on the  the  CP violation sensitivity using different channels.
   We assess in a comprehensive manner the sensitivity to CP violation offered by present and future generation long baseline experiments : \ttok\,, \nova\,, \dune\, and \ttohk\, when the NSI effects are turned on.

The paper is organised as follows. Sec.~\ref{sec:framework} gives the framework for the present work. Sec~\ref{framework_a} comprises of a brief introduction to NSI in propagation and how the genuine and fake CP violating effects can arise due to NC NSI terms which are constrained.  We then discuss the CP dependence of the probabilities 
$\pme$ and $P_{\mu\mu}$ in Sec.~\ref{framework_b}. In Sec.~\ref{framework_c}, we give our analysis procedure using the CP dependence of probabilities in Sec.~\ref{framework_b}.
We then go on to describe our results in Sec.~\ref{results} where we show how CP sensitivity at \dune\, gets affected due to individual and collective NSI terms (Sec.~\ref{results_a}). 
We also show dependence on true values of standard oscillation parameters in Sec.~\ref{results_b} and compare the results obtained at \dune\, with other long baseline experiments in 
Sec.~\ref{results_c}. The impact of NSI on the CP fraction is shown as a 
function of exposure and baseline in Sec.~\ref{results_d} and \ref{results_e}. A discussion of CP violation sensitivity when the source of CP violation is known is given in Sec.~\ref{results_f}. Finally, the ability of long baseline experiments to measure the CP phases is discussed in Sec.~\ref{results_g}.
We conclude with a discussion  in Sec.~\ref{sec:conclude}.

\section{Framework}
\label{sec:framework}
\subsection{Genuine and fake CP violation due to Earth matter effects - SI and NSI}
\label{framework_a}

It is well-known that the three neutrino flavor states can be mapped to a three-level quantum system with distinct energy eigenvalues, $E_i = p + m^2_i /2p$ in the ultrarelativistic limit in vacuum along 
with the assumption of equal fixed momenta (or energy). 
In the presence of matter, the relativistic dispersion relation $E_i = f(p,m_i)$ gets modified due 
to the neutrino matter interactions during propagation. 
The effective Lagrangian describing the NC type neutrino NSI of the
type $(V-A)(V \pm A)$ is given by
\begin{equation}
\label{nsilag}
{\cal L}_{NSI} = -2 \sqrt 2 G_F \varepsilon_{\alpha \beta}^{f\, C} ~ [\bar \nu_\alpha \gamma^\mu P_L \nu_\beta] ~[\bar f \gamma_\mu P_C f]~,
\end{equation}
where $G_F$ is the Fermi constant, $\nu_{\alpha},\nu_{\beta}$ are
neutrinos of different flavours, and $f$ is a first generation SM
fermion ($e,u,d$)~\footnote{Coherence requires that the flavour of the
background fermion ($f$) is preserved in the interaction.  Second or
third generation fermions do not affect oscillation experiments since
matter does not contain them.}.  The chiral projection operators are
given by $P_L = (1 - \gamma_5)/2$ and $P_C=(1\pm \gamma
_5)/2$. In general, NSI terms can be complex. It should be noted that charged current (CC) NSI would only affect neutrino production and detection as opposed to NC NSI and hence as far as propagation of neutrinos is concerned, only the NC NSI is relevant. In the effective
 \sch equation for neutrino propagation, the effective Hamiltonian in flavour basis is given by
 \bea
 \label{hexpand} 
 {\mathcal
H}^{}_{\mathrm{f}} &=&   {\mathcal
H}^{}_{\mathrm{v} } +  {\mathcal
H}^{}_{\mathrm{SI} } +  {\mathcal
H}^{}_{\mathrm{NSI}} 
\nonumber 
\\
&
=&\lambda \left\{ {\mathcal U} \left(
\begin{array}{ccc}
0   &  &  \\  &  r_\lambda &   \\ 
 &  & 1 \\
\end{array} 
\right) {\mathcal U}^\dagger  + r_A   \left(
\begin{array}{ccc}
1  & 0 & 0 \\
0 &  0 & 0  \\ 
0 & 0 & 0 \\ 
\end{array} 
\right)  +
 {r_A}   \left(
\begin{array}{ccc}
\epsilon_{ee}  & \epsilon_{e \mu}  & 
\epsilon_{e \tau}  \\ {\epsilon_{e\mu} }^ \star & 
\epsilon_{\mu \mu} &   \epsilon_{\mu \tau} \\ 
{\epsilon_{e \tau}}^\star & {\epsilon_{\mu \tau}}^\star 
& \epsilon_{\tau \tau}\\
\end{array} 
\right) \right\}  \ ,
 \eea 
where we have used the following ratios
\begin{equation}
\lambda \equiv \frac{\delta m^2_{31}}{2 E}  \quad \quad ; \quad \quad
r_{\lambda} \equiv \frac{\delta m^2_{21}}{\delta m^2_{31}} \quad \quad ; 
\quad \quad r_{A} \equiv \frac{A (x)}{\delta m^2_{31}} \ .
\label{dimless}
\end{equation}
and  the standard CC potential due to
the coherent forward scattering of neutrinos is given by 
 $A (x)= 2 \sqrt{2}  	E G_F n_e (x)$  where  $n_e$ is the electron
number density. 
${\mathcal
  U}$
is the three flavour neutrino mixing matrix and is responsible for diagonalizing the vacuum part of the Hamiltonian. 
It is parameterized by three angles $\theta_{12},\theta_{23},\theta_{13}$ and 
one  phase  $\delta$
\bea 
{\cal U} (\{\theta_{ij}\},\delta) &\equiv& 
{\cal U}_{23}  (\theta_{23})  \cdot 
\, {\cal W}_{13} (\theta_{13},\delta) \cdot 
\,{\cal U}_{12} (\theta_{12})\eea
 with
  ${\cal W}_{13} = {\cal U}_\delta~ {\cal U}_{13}~ {\cal
    U}_\delta^\dagger$ and ${\cal U}_\delta = {\mathrm{diag}}
  \{1,1,\exp{(i \delta)}\}$] ~\footnote{In the general case of n flavors the leptonic mixing matrix 
$U_{\alpha i}$ depends on $(n-1)(n-2)/2$ Dirac-type 
CP-violating  phases. 
If the neutrinos are Majorana particles, there are $(n-1)$ additional, so called Majorana-type CP-violating phases. } 
In the commonly used Pontecorvo-Maki-Nakagawa-Sakata (PMNS) parametrization~\cite{Beringer:1900zz}, ${\cal U}$ is given by
\bea
{\mathcal U}^{} &=& \left(
\begin{array}{ccc}
1   & 0 & 0 \\  0 & c_{23}  & s_{23}   \\ 
 0 & -s_{23} & c_{23} \\
\end{array} 
\right)   
  \left(
\begin{array}{ccc}
c_{13}  &  0 &  s_{13} e^{- i \delta}\\ 0 & 1   &  0 \\ 
-s_{13} e^{i \delta} & 0 & c_{13} \\
\end{array} 
\right)  \left(
\begin{array}{ccc}
c_{12}  & s_{12} & 0 \\ 
-s_{12} & c_{12} &  0 \\ 0 &  0 & 1  \\ 
\end{array} 
\right)  \ ,
\label{u}
 \eea 
where $s_{ij}=\sin {\theta_{ij}}, c_{ij}=\cos \theta_{ij}$.  If neutrinos are Majorana particles, there can be 
two additional Majorana-type phases in the three flavour case but they are of no consequence 
 in neutrino oscillations.  For the SI case, we note that there is only one parameter, the Dirac CP phase $\delta$
 that is  responsible for genuine CP violation while SI with Earth matter introduces additional 
  fake CP effects due to the fact that matter is CP asymmetric. This makes it challenging to isolate the value of   genuine CP violating phase $\delta$ in SI case from the fake effects  and there are several suggestions to tackle the problem~\cite{Arafune:1997hd,Ohlsson:2013ip}.  The geometric visualization of 
CP conservation and CP violation for the two flavour neutrino case was demonstrated in~\cite{Mehta:2009ea,Mehta:2009xm}.
 
 For the NSI case,  the ${\epsilon}_{\alpha \beta} \, (\equiv |\epsilon _{\alpha \beta}|
\, e^{i \phi_{\alpha\beta}})$ are complex NSI parameters which appear in ${\cal H}_{NSI}$. The diagonal NSI parameters  are real due to the hermiticity of the Hamiltonian. 
 In total, there are four phases appearing in the ${\cal H}_{\rm f}$ - one is $\delta$ and the other three are $\eemp,\eetp,\emtp$.  The total number of phases is four due to the fact 
 that once we have redefined the phases of the  lepton and neutrino wavefunctions to get ${\cal U}$ in form Eq.~\ref{u} the basis of neutrino flavor states in defined fully. The matrix that diagonalizes the NSI 
  part of the Hamiltonian then would require three angles and six phases out of which three  are Majorana type and appear as diagonal matrix. So, we are left with 
  three additional phases that are relevant for us. It may be further possible to reduce the number of phases in the limiting cases such as 
 when $\delta m^2_{21}  \to 0$ or $\theta_{12}  \to 0$ as a consequence of 
 the phase reduction theorem of Kikuchi et al.~\cite{Kikuchi:2008vq}.

 For long baseline neutrino experiments, $\sin^2 (\lambda L/2) \simeq {\cal O} (1)$ which gives 
\begin{eqnarray}
\frac{\lambda L}{2} &\simeq& 1.57~ \left[ \frac{\delta m^2_{31}}{2.5 \times 10^{-3}~eV^2} \frac{2.5~GeV}{E} \frac{L}{1300~km} \right]     ~ \quad \quad {\rm{for ~\dune}}, 
\end{eqnarray}
for the first oscillation maximum (minimum) in the appearance (disappearance) channel. We note that $E=1.5~GeV, L=810~km$ for \nova \, and $E=0.6~GeV, L=295~km$ for \ttok\, (and also \ttohk) also lead to $\lambda L \sim \pi$.
Also, $r_A L \sim {\cal O} (1)$ for the range of  the $E$ and $L$ values considered here.

It is interesting to note that matter (or propagation) NSI obey unitarity (while source and detector NSI do not) so effectively we still 
have an overall unitary matrix that diagonalises the effective Hamiltonian in presence of matter NSI and obeys
\begin{equation}
\sum_i \hat {\cal U}_{\alpha i} ~\hat {\cal U}_{\beta i}^\star   = \delta_{\alpha \beta}\end{equation}
where $\hat {\cal{ U}}$ is the unitary matrix that diagonalizes the Hamiltonian in Eq.~\ref{hexpand}.
\bea
 {\mathcal
H}^{}_{\mathrm{d}} &=&  \hat{\mathcal U} ^\dagger ~{\cal H}_{\rm f}~ \hat {\mathcal U}   \ ,
 \eea 
 where the elements ${\cal H}_{\rm d\, ii}$ are the eigenvalues of $H_{\rm f}$.

As far as the constraints on NC NSI parameters are concerned, 
we refer the reader to Ref.~\cite{Chatterjee:2014gxa,Ohlsson:2012kf} 
for more details. After taking the constraints from neutrino experiments into account, the NSI parameters are constrained as follows
 \begin{eqnarray}
 |\varepsilon_{\alpha\beta}|
 \;<\;
  \left( \begin{array}{ccc}
4.2  &
0.3 & 
0.5 \\
0.3 & 0.068 & 0.04 \\
0.5  &
0.04 &
0.15 \\
  \end{array} \right) \ .\label{tinynsi}
\end{eqnarray} 
The NSI phases are unconstrained and can lie the allowed range, $\varphi_{\alpha\beta} \in (-\pi,\pi)$ (see Table.~\ref{tab:parameters}).

All the plots presented in this paper are obtained by using General Long baseline Experiment Simulator (GLoBES) and related software~\cite{Huber:2004ka,Kopp:2006wp,Huber:2007ji,Kopp:2007ne} which numerically 
solves  the full three flavour neutrino propagation equations using the PREM~\cite{Dziewonski:1981xy} 
\begin{figure}[th!]
\centering
\includegraphics[width=\textwidth]
{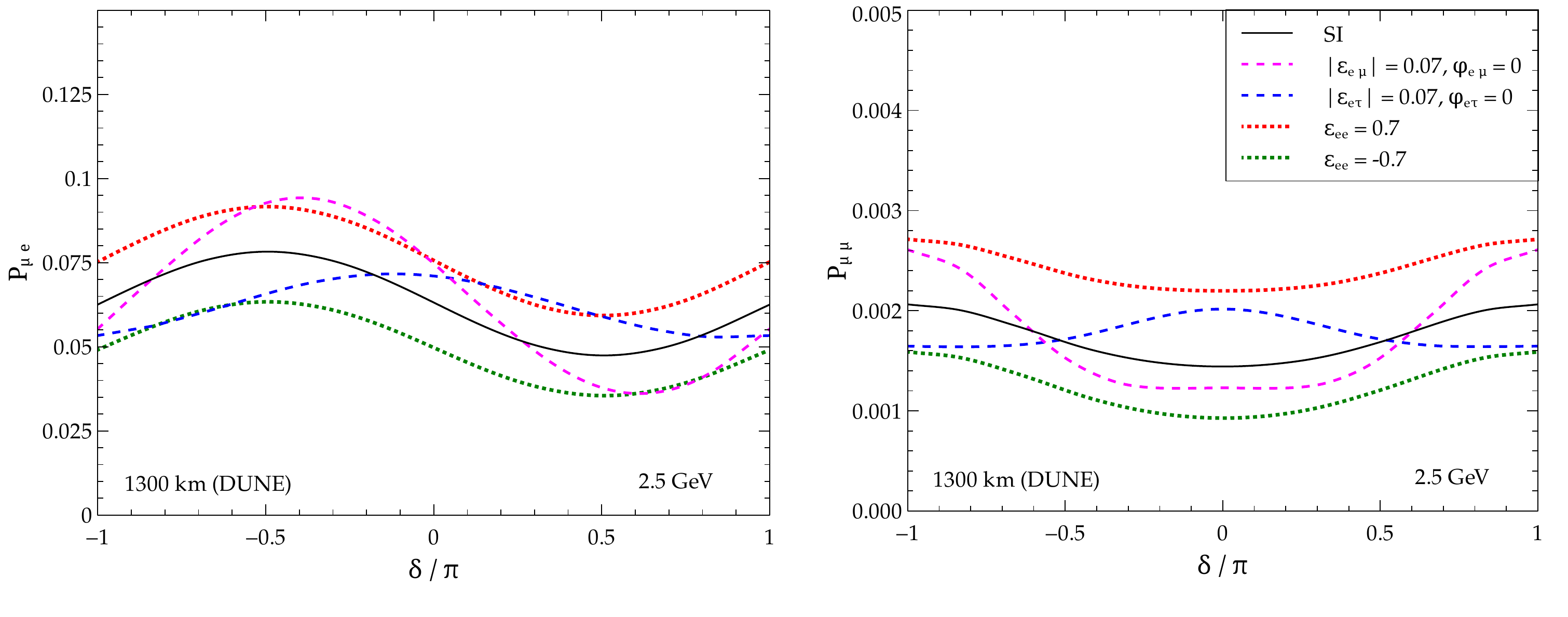}
\caption{\footnotesize{Effect of individual NSI  terms  in the $P_{\mu e}$ and $P_{\mu\mu}$ as a function of $\delta$ for $E=2.5$ GeV and $L=1300$ km. The solid black curve represents SI case while the dashed (dotted) curves  represent the case of off-diagonal (diagonal) NSI parameters. The NSI phases  $\phi_{e\mu}$ and $ \phi_{e\tau}$ are set to zero.
}}
\label{fig:prob_app_disapp_ind}
\end{figure}
density profile of the Earth\footnote{We use the matter density as given by PREM model.
 In principle, we can allow for uncertainty in the Earth matter density in our calculations but  it would not impact our results drastically~\cite{Gandhi:2004md,Gandhi:2004bj}.},
 and the latest values of the neutrino parameters as obtained from global
fits~\cite{GonzalezGarcia:2012sz,Capozzi:2013csa,Forero:2014bxa}. Unless stated otherwise, we assume NH as the true hierarchy in all the plots.

\subsection{CP phase dependence in $P_{\mu e}$ and $P_{\mu\mu}$}
\label{framework_b}

 We consider  appearance ($\nu_\mu \to \nu_e$) and disappearance ($\nu_\mu \to \nu_\mu$) channels that are relevant in the context of accelerator-based neutrino oscillation experiments considered in the present work.  Rather than delving  into the detailed expressions, we note that~\cite{Kimura:2002hb,Kimura:2002wd,Ohlsson:2013ip} that the oscillation probabilities for different channels can be expressed in terms of the CP-even and CP-odd  terms both in case of vacuum and matter with SI~\footnote{no extra phase, with suitable redefinition of coefficients} as 
\begin{enumerate}
\item $\nu_\mu \to \nu_ e$ and $\bar\nu_\mu \to \bar \nu_ e$ : 
\bea
P_{\mu e} &=& a_{\mu e} + b_{\mu e} \sin \delta + c_{\mu e} \cos \delta \nonumber\\
P_{\bar\mu \bar e} &=& \bar a_{\mu e} - \bar b_{\mu e} \sin \delta + \bar c_{\mu e} \cos \delta
\label{pap}
\eea
 $\delta \to -\delta$ for antineutrinos and the coefficients can be found in Ref.~\cite{Kimura:2002wd}. Thus, $P_{\mu e}$ contains linear polynomials of $\sin \delta$ and $\cos \delta$. 
\item $\nu_\mu \to \nu_\mu$ and $\bar \nu_\mu \to \bar \nu_\mu$ : 
\bea
P_{\mu \mu} &\simeq& a_{\mu\mu} + c_{\mu\mu} \cos \delta \nonumber\\
P_{\bar\mu \bar \mu} &\simeq& \bar a_{\mu\mu} + \bar c_{\mu\mu} \cos \delta 
\label{pdi}
\eea

\begin{figure}[htb]
\centering
\includegraphics[width=0.8\textwidth]
{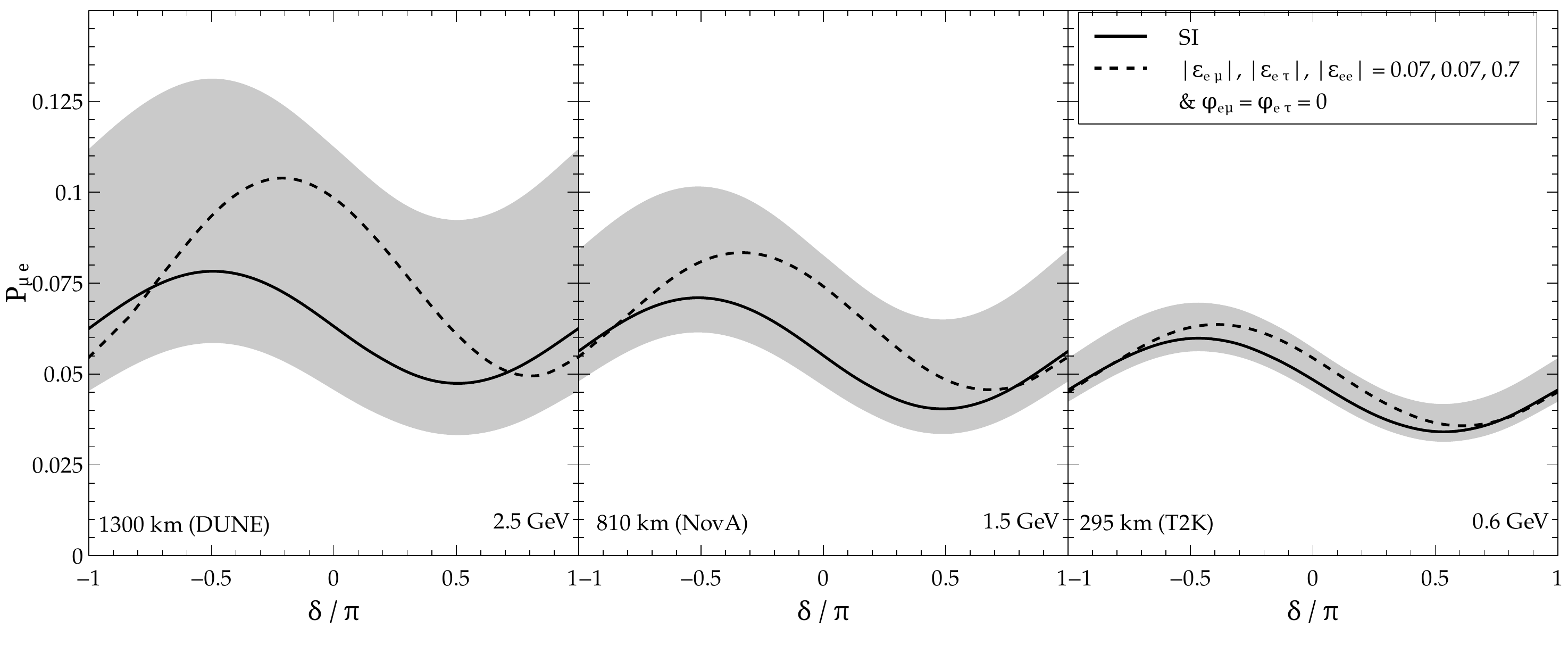}
\includegraphics[width=0.8\textwidth]
{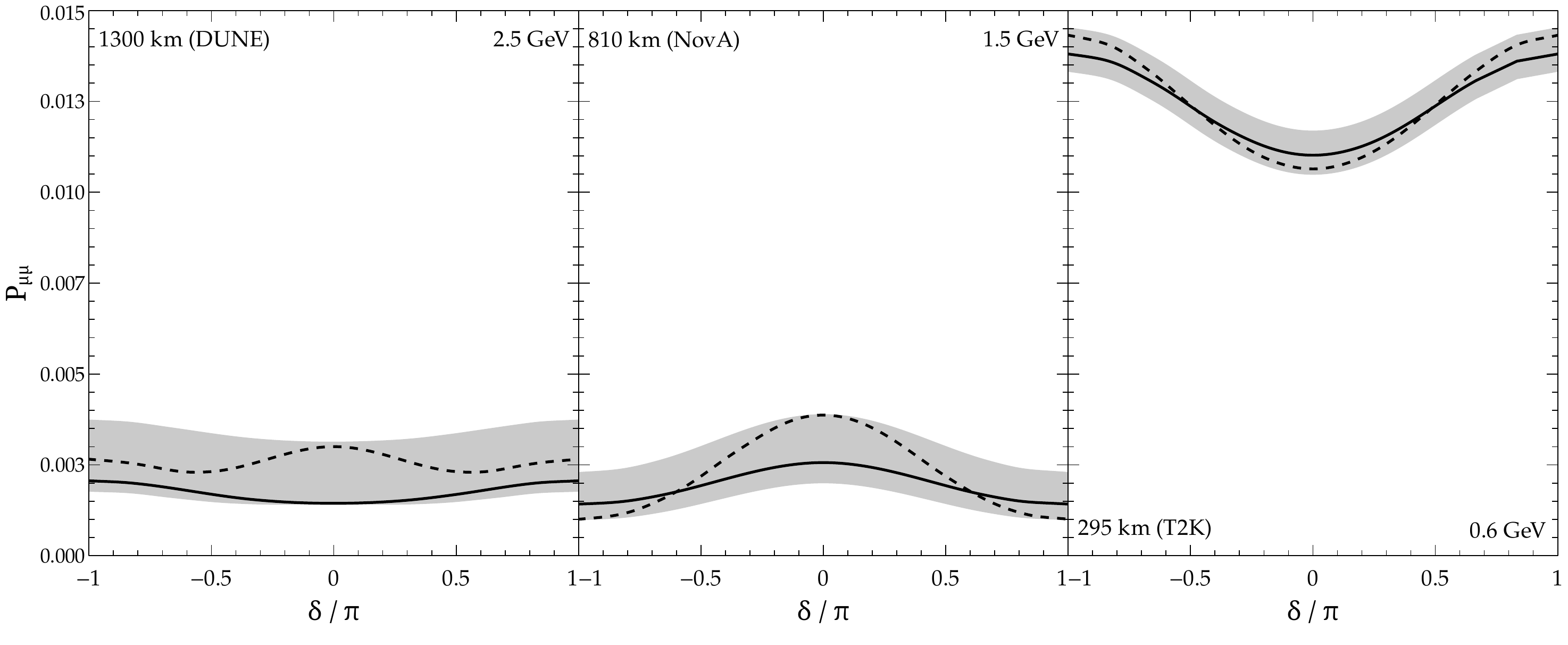}
\caption{\footnotesize{Combined effect of three NSI terms ($\eem,\eet,\eee$) in the 
electron appearance and muon disappearance  probability as a function of $\delta$ (for fixed $E$ and $L$ for \dune, \nova\, and \ttok). The solid black curve represents  SI   while the dashed black curve represents  NSI for the particular choice of absolute value of NSI parameters as mentioned in the legend. The grey band shows the spread when in addition the NSI phases are varied in the
  allowed range \ie,  
 $\phi_{e\mu},\phi_{e\tau} \in [-\pi,\pi]$. 
 }}
\label{fig:prob_app_disapp_com}
\end{figure}
 
where the $\sin \delta$ term is absent in this case. In addition to the linear polynomials of  $\cos\delta$ in this case, there are quadratic terms such as $\cos 2 \delta$ (and $\sin 2\delta$) in $P_{\mu\mu}$ for the case of constant or symmetric (asymmetric) matter density profile, but the coefficient of such terms are  small in comparison to $a_{\mu\mu}$ and $c_{\mu\mu}$ which is why we do not explicitly mention those here. 
For antineutrinos, $\delta \to -\delta$ and  the coefficients in vacuum and normal matter can be found in Ref.~\cite{Kimura:2002wd}.
\end{enumerate}
It is interesting to note that in presence of matter with SI, the form of Eqs.~\ref{pap}-\ref{pdi}  remain intact with the coefficients  suitably redefined to account for their dependence on density of Earth matter. The CP odd and even terms in Eq.~\ref{pap} and \ref{pdi} serve as useful guide to measure effects due to CP violation.

Let us now discuss the case of NSI which is different from SI in the sense that it not only introduces  SI matter-like fake CP violating effects arising from the moduli of the NSI terms but also additional genuine CP phases over and above SI phase ($\delta$). Of course, the  genuine and fake CP violating effects are inter-related.  
  The argument of Kimura et al.~\cite{Kimura:2002hb,Kimura:2002wd} was generalised to the case of NSI~\cite{Yasuda:2007jp,Meloni:2009ia}.  In \cite{Yasuda:2007jp} (for non-zero values of $\eet,\eee,\ett$), it was shown that the CP dependence of  probability given by Eq.~\ref{pap} and \ref{pdi} remains intact even in presence of NSI   as long as we make appropriate replacement for the effective CP violating phase, $\delta \to \hat \delta$. 
   
In Fig.~\ref{fig:prob_app_disapp_ind}, the impact of individual  NSI terms on $P_{\mu e}$ and $P_{\mu\mu}$ is shown for the baseline corresponding to \dune\, ($L=1300$ km) at a fixed value of energy $E=2.5$ GeV. 
 A striking feature can be clearly seen if we compare the nature of plots on left side versus the right  side.  We can simplify Eq.~\ref{pap} to
\bea
P_{\mu e} &=& a_{\mu e} - x'_{\mu e} \sin \lambda L/2 \sin \delta + x'_{\mu e} \cos \lambda L/2 \cos \delta \nonumber\\
 &=& a_{\mu e} + x'_{\mu e} \cos (\lambda L/2 + \delta) 
\label{p_piby2}
\eea
where $x'_{\mu e}$ can be found in Ref.~\cite{Barger:2001yr} for SI.
We note that the (first) peak condition for $P_{\mu e}$ is given by $\lambda L/2 \simeq \pi/2$ and therefore for a given $\delta$, the peak (dip) of  $P_{\mu e}$ is shifted by an amount $\pi/2$  w.r.t. $P_{\mu\mu}$ (see Eq.~\ref{pdi} and \ref{p_piby2}).
For SI, Eq.~\ref{pap} leads to $P_{\mu e} (0) = a_{\mu e}+c_{\mu e} $ and $P_{\mu e} (\pm \pi) = a_{\mu e}-c_{\mu e}$. From the plot, we see that  $P_{\mu e} (0) \sim P_{\mu e} (\pm \pi)$ and this implies that $c_{\mu e} \simeq 0$. Note also that the maxima/minima will be at $\delta = \pm \pi/2$  from  Eq.~\ref{pap}. If we keep the relevant NSI phases to zero, the  dashed (dotted) curves corresponding to NSI can go on either side of the solid curve for SI.   For SI, Eq.~\ref{pdi} leads to $P_{\mu \mu} 
(0) = a_{\mu \mu}+c_{\mu \mu} $ and $P_{\mu \mu} (\pm \pi) = a_{\mu \mu}-c_{\mu \mu}$. 
 Note also that the maxima/minima will be at $\delta = 0$ or $\pi$  from  Eq.~\ref{pdi}. For the diagonal parameter $\varepsilon_{ee}$, for both $P_{\mu e}$ and $P_{\mu\mu}$, the effect is like a uniform enhancement (reduction) of the probability values from the SI case depending upon the sign of $\varepsilon_{ee}$.

In Fig.~\ref{fig:prob_app_disapp_com}, the  collective impact of NSI terms is shown for three 
different experiments at different fixed energies relevant to those experiments.  
The largest  effect of NSI terms can be seen for $P_{\mu e}$ and for \dune\, and it diminishes as we go to \ttok. For $P_{\mu\mu}$, the effect is  similar for all the three experiments so the baseline does not seem to play much role here.

\begin{table}[h]
\centering
\begin{tabular}{ |l l l|}
\hline
&&\\
Parameter & True value & Marginalisation range  \\
&&\\
\hline
&&\\
{\sl{SI}} &&\\
&&\\
$\theta_{12}$ [deg] & 33.5  &  - \\
$\theta_{13}$ [deg] & 8.5 & -\\
$\theta_{23}$ [deg] & 45  & - \\
$\delta m^2_{21}$ [$eV^2$]  & $7.5 \times 10^{-5}$ & - \\
$\delta m^2_{31}$ (NH) [$eV^2$] & $+2.45 \times 10^{-3}$ & - \\
$\delta m^2_{31}$ (IH) [$eV^2$] & $-2.46 \times 10^{-3}$ & - \\
$\delta$ & $[-\pi:\pi]$ & $0, \pi$\\
&&\\
\hline
&&\\
{\sl{NSI}} &&\\
&&\\
$\eee$ & $ 0.1, 0.4, 0.7$ & $[0 : 1.00]$ \\
$\emm$ & $ 0.05$ & $[0 : 0.06]$ \\
$\ett$ & $ 0.04, 0.08, 0.12$ & $ [0 : 0.15]$ \\
$|\eem|$ & $0.01,0.04,0.07$ & $[0:0.10]$ \\
$|\eet|$ & $0.01,0.04,0.07$ & $[0:0.10]$ \\
$|\emt|$ & $0.01,0.04$ & $[0:0.04]$\\
$\varphi_{e\mu}$ & $[-\pi:\pi]$& $0, \pi$ \\
$\varphi_{e\tau}$ & $[-\pi:\pi]$& $0, \pi$ \\
$\varphi_{\mu\tau}$ & $[-\pi:\pi]$& $0, \pi$ \\
\hline
\end{tabular}
\caption{\label{tab:parameters}
 SI and NSI parameters used in our study. For latest global fit to neutrino data see~\cite{Gonzalez-Garcia:2014bfa}.}
\end{table}

\begin{table}[ht]
\centering
\begin{tabular}{ |l| l l |ll| }
\hline
Detector details & \multicolumn{2}{c |}{Normalisation error} & \multicolumn{2}{c |}{Energy calibration error} \\
                                         \cline{2-5}
                                        & Signal & Background       & Signal & Background                              
                                        \\ 
\hline
\dune 
&&&&\\
 Runtime (yr) = 5 $\nu$ + 5 $\bar \nu$ 
  & $\nu_e : 5\%$  & $\nu_e : 10\%$ &   $\nu_e : 2\%$ & $\nu_e : 10\%$\\
35 kton, LArTPC & & &&\\
$\epsilon_{app}=80\%$, $\epsilon_{dis}=85\%$
& $\nu_\mu : 5\%$ & $\nu_\mu : 10\%$ & $\nu_\mu : 5\%$ & $\nu_\mu : 10\%$\\
{{$R_\mu=0.20/{\sqrt E}$, $R_e=0.15/{\sqrt E}$}} &&&&\\
\hline
\nova &&&&\\
 Runtime (yr) = 3 $\nu$ + 3 $\bar \nu$ 
&
 $\nu_e : 5\%$  & $\nu_e : 10\%$ &   $\nu_e : 0.01\%$ & $\nu_e : 0.01\%$\\
14 kton, TASD  & & &&\\
$\epsilon_{app}=55\%$, $\epsilon_{dis}=85\%$
& $\nu_\mu : 2.5\%$ & $\nu_\mu : 10\%$ & $\nu_\mu : 0.01\%$ & $\nu_\mu : 0.01\%$\\
{{$R_\mu=0.06/{\sqrt E}$, $R_e=0.085/{\sqrt E}$}} 
&&&&\\
\hline
T2K 
&&&&\\
 Runtime (yr) = 3 $\nu$ + 3 $\bar \nu$  %
 & $\nu_e : 5\%$  & $\nu_e : 5\%$ &   $\nu_e : 0.01\%$ & $\nu_e : 0.01\%$\\
 22.5 kton, WC & & &&\\
$\epsilon_{app}=50\%$, $\epsilon_{dis}=90\%$
& $\nu_\mu : 2.5\%$ & $\nu_\mu : 20\%$ & $\nu_\mu : 0.01\%$ & $\nu_\mu : 0.01\%$\\
{{$R_\mu=0.085/{\sqrt E}$, $R_e=0.085/{\sqrt E}$}} &&&&\\
\hline
\ttohk
&&&&\\
Runtime (yr) = 1 $\nu$ + 3 $\bar \nu$
  & $\nu_e : 5\%$  & $\nu_e : 5\%$ &   $\nu_e : 0.01\%$ & $\nu_e : 0.01\%$\\
560 kton, WC & & &&\\
$\epsilon_{app}=50\%$, $\epsilon_{dis}=90\%$ & $\nu_\mu : 2.5\%$ & $\nu_\mu : 20\%$ & $\nu_\mu : 0.01\%$ & $\nu_\mu : 0.01\%$\\
{{$R_\mu=0.085/{\sqrt E}$, $R_e=0.085/{\sqrt E}$}} 
&&&&\\
\hline
\end{tabular}
\caption{\label{tab:sys} 
Detector configuration, efficiencies, resolutions and systematic uncertainties for \dune, \nova, \ttok, \ttohk.  }
\end{table}

  \subsection{Analysis procedure}
  \label{framework_c}

In order to obtain the sensitivity to CP violation we need to ask the following question - 
what is the sensitivity with which a particular experiment can discriminate between CP conserving ($ 0,\pi$) and CP violating values ($\neq 0,\pi$) of the Dirac CP phase $\delta$.
In the standard scenario, there is only one CP phase in the neutrino oscillation formalism. However when we consider NSI, naturally more parameters in the form of moduli and phases of NSI parameters 
enter the oscillation  formalism which lead to genuine and fake CP violating effects  as described above.  

For the purpose of understanding the gross features in the plots, we give below the statistical definition of $\chi^2$ for CP violation sensitivity,
\begin{equation}
\label{chisq}
\chi^2 \equiv  \min_{\delta,|\varepsilon|,\varphi}  \sum_{i=1}^{x}  \sum_{j}^{2} 
 \frac{\left[N_{true}^{i,j}(\delta,|\varepsilon|,\varphi) - N_{test}^{i,j} (\delta=0,\pi ; ~|\varepsilon| ~{\rm{range}};~ \varphi = 0,\pi)\right]^2 }{N_{true}^{i,j} (\delta,|\varepsilon|,\varphi)}~,
\end{equation}
where $N_{true}^{i,j}$ and $N_{test}^{i,j}$ are the number of true and test events in the $\{i,j\}$-th bin respectively~\footnote{$N_\sigma = \sqrt{\Delta \chi^2}$. 
$\Delta \chi^2 = \chi^2$ as we have not included any fluctuations 
in simulated data. 
This is the Pearson's definition of $\chi^2$~\cite{Qian:2012zn}. For large sample size, the other 
 definition using log-likelihood also yields similar results.} . The NSI parameters are expressed in terms of moduli, $|\varepsilon| \equiv \{ |\varepsilon_{\alpha\beta}| ; \alpha,\beta = e,\mu,\tau\}$ and phases, $\varphi \equiv \{\varphi_{\alpha\beta} ; \alpha,\beta = e,\mu,\tau \}$. The marginalisation range of NSI parameters ($|\varepsilon| ~{\rm{range}}$ and $\varphi$)  and SI parameters ($\delta$) is given in Table~\ref{tab:parameters}. 
To determine the $ \chi^2$ that represents a particular experiment's  sensitivity to the presence of CP violating effects, the test value of phases 
($\delta,\varphi_{e\mu},\varphi_{e\tau}$) is assumed to be $0$ or $\pi$ and the  $ \chi^2$  for any true value of phase ($\delta,\varphi_{e\mu},\varphi_{e\tau}$) in the full range of $[-\pi,\pi]$ is computed\footnote{$\delta,\varphi_{e\mu},\varphi_{e\tau}$ 
can take $0,\pi$ and there there are eight possibilities for CP conservation in NSI case as opposed to two in SI case.}. While the variation corresponding to the true value of $\delta$ is depicted along the $x-$axis, the variation  of the true values of $\varphi_{e\mu},\varphi_{e\tau}$ (with in the allowed range) lead to the vertical width of the grey bands which show the maximum variation in the $\chi^2$ for each value of $\delta$ (true)\footnote{The true values of $\varphi_{e\mu}$ and $\varphi_{e\tau}$ that lead to maximum and minimum $\chi^2$ are in general not the same for each $\delta$ (true). }.

We do not marginalise over the standard oscillation parameters except $\delta$ whose true value is unknown. 
As we are investigating the role of NSI in the present study, we marginalize over the allowed range of moduli and phases of the relevant NSI parameters. Our choice of range of NSI parameters is consistent with 
the  existing constraints (Eq.~\ref{tinynsi}). 

The indices $i,j$ correspond to energy bins  ($i=1 \to x$, the number of bins depends upon the particular experiment - for \dune, there are $x=39$ bins of width 250 MeV in $0.5-10$ GeV, for \ttok\, and \ttohk, there are $x=20$ bins of width 40 MeV  in $0.4-1.2$~GeV, for \nova, there are $x=28$ bins of width 125 MeV in $0.5-4$ GeV) and the type of neutrinos i.e. neutrino or antineutrino ($j = 1 \to 2$).  As one would expect, the discovery potential of a given experiment vanishes for the CP conserving case ($\delta,\varphi = 0$ or $\pi$). At the values of $\delta$ corresponding to maximum CP violation i.e. $\delta=\pm\pi/2$, the discovery potential reaches a maximum. So, there is a double peak structure in the sensitivity plot.  This can also be understood from the probability curves for SI in Figs.~\ref{fig:prob_app_disapp_ind}-\ref{fig:prob_app_disapp_com}. The appearance probability has a maxima and minima at $\delta = \pm \pi/2$. 

For  sake of clarity, we have retained only statistical effects and ignored systematic uncertainties and priors  in the above expression (see~\cite{Barger:2014dfa} for the full expression of $\chi^2$ including systematics and priors). 
However, in our analysis, we have marginalised over systematic uncertainties (see Table~\ref{tab:sys}) but  assumed that the standard oscillation parameters are known with infinite precision i.e. we have not  included any priors.  

The theoretically expected differential event rate is given by~\cite{Bass:2013vcg},  
\bea
\label{eq:eventrate}
\frac{dN^{app}_{\nu_e} (E,L)}{dE}  &=& N_{target}  \times 
 \Phi_{\nu_\mu} (E,L)  \times P_{\mu e} (E,L) \times \sigma_{\nu_e} (E) ~,%
 \eea where
  $N_{target}$ is the number of target nucleons per kiloton of detector fiducial volume, $N_{target}=6.022 \times 10^{32}~N/kt$. 
  $P_{\mu e}  (E,L)$ is the appearance probability for $\nu_\mu \to \nu_e$ in matter, $\Phi_{\nu_\mu} (E,L)$ is the flux of $\nu_\mu$, 
 $\sigma _{\nu_e} (E)$ is the  CC cross section of $\nu_e$ given by 
 \begin{equation}
 \sigma_{\nu_e} = 0.67 \times 10^{-42} (m^2/GeV/N) \times E~, \quad {\rm {for}} \quad E > 0.5~ {\rm{GeV}}
 \end{equation}
 For 
  the disappearance channel, $P_{\mu e} $ is to be replaced by $P_{\mu \mu}$ and $\sigma_{\nu_e} \to \sigma_{\nu_\mu}$.  Note that 
 $\sigma_{\nu_\mu } \sim \sigma_{\nu_e}$ for the considered energy range.
  For antineutrinos, $\nu_\mu \to \bar\nu_\mu$ and $\nu_e \to \bar\nu_e$ and $P_{\mu e} \to \bar P _{\mu e}$. 
   
 The $\chi^2$ for appearance  channel is obtained by adding the neutrino ($\nu_\mu \to \nu_e$) and antineutrino ($\bar \nu_\mu \to \bar\nu_e $) contributions\footnote{Note that the factors such as $N_{target}$, $\Phi_{\nu_\mu}$ ($\bar\Phi_{\nu_\mu}$) and $\sigma_{\nu_\mu/\nu_e}$ ($\sigma_{\bar\nu_\mu/\bar\nu_e}$) will also be present, but they are independent of the CP phase and hence omitted in the discussion that follows.}, which gives 
\begin{eqnarray}
\chi^2_{app} &=& \chi^2_{\nu_\mu \to \nu_e} + \chi^2_{\bar \nu_\mu \to \bar \nu_e } ~,
\nonumber \\
&\propto& \min_{0,\pi} \left\{
[P^{true}_{\mu e} - P ^{test}_{\mu e}]^2 +	[\bar P^{true}_{\mu e} - 
\bar P^{test}_{\mu  e}]^2 \right\}~,
\nonumber\\
&=& 
\min_{0,\pi} \bigg\{ \left[b_{\mu e} \sin {\delta}_{true} + c_{\mu e} \cos \delta _{true} - 
c_{\mu e} \cos \delta \vert_{0,\pi} \right]^2 
\nonumber\\
&  & + \left[-\bar b_{\mu e} \sin {\delta}_{true} + \bar c _{\mu e} \cos \delta _{true} - 
\bar c_{\mu e} \cos \delta \vert_{0,\pi}\right]^2 \bigg\}~.
\label{eq:chi_app}
\end{eqnarray}
Here the $\sin \delta_{test}$ term  vanishes while $\cos \delta_{test}$ does not for the two 
 CP conserving values $\delta=0,\pi$.  
  Similarly, the $\chi^2$ for disappearance  channel is obtained by adding the neutrino ($\nu_\mu \to \nu_\mu$) and antineutrino ($\bar \nu_\mu \to \bar \nu_\mu$) contributions,
\begin{eqnarray}
\chi^2_{dis} &=& \chi^2_{\nu_\mu \to \nu_\mu} + \chi^2_{\bar \nu_\mu \to \bar \nu_\mu }~, 
\nonumber \\
&\propto& 
[P^{true}_{\mu \mu} - P ^{test}_{\mu \mu}]^2 - 	[P^{true}_{\bar \mu \bar \mu} - 
P^{test}_{\bar \mu \bar \mu}]^2~,
\nonumber \\
&=&\min_{0,\pi} \bigg\{ \left[ c_{\mu\mu} \cos \delta _{true}  - 
c_{\mu\mu} \cos \delta\vert_{0,\pi} \right]^2 
 + \left[ \bar c_{\mu\mu} \cos \delta _{true}  - 
\bar c_{\mu\mu} \cos \delta \vert_{0,\pi}  \right]^2 \bigg\}~.
\end{eqnarray}
Note that both $\chi^2_{app}$ and $\chi^2_{dis}$ depend on  $\cos \delta_{true}$ while $\chi^2_{app}$ also depends on
$\sin \delta_{true}$. 
 The presence of $\sin \delta_{true}$ term in the $\chi^2_{app}$ ensures that 
the  appearance channel contributes dominantly to the CP violation sensitivity.
The total $\chi^2$ when appearance and disappearance channels are combined is given by
\begin{eqnarray}
\chi^2_{tot} &\propto&\min_{0,\pi} \bigg\{  
\left[b_{\mu e} \sin {\delta}_{true} + c_{\mu e} \cos \delta _{true} - 
c_{\mu e} \cos \delta\vert_{0,\pi}\right]^2 
\nonumber\\ &&
 ~+ \left[-\bar b_{\mu e} \sin {\delta}_{true} + \bar c_{\mu e} \cos \delta _{true} - 
\bar c_{\mu e} \cos \delta\vert_{0,\pi}\right]^2 \nonumber\\
&& ~+
\left[ c_{\mu\mu} \cos \delta _{true}  - 
c_{\mu\mu} \cos \delta\vert_{0,\pi}   \right]^2 
+ \left[ \bar c_{\mu\mu}
 \cos \delta _{true}  - 
\bar  c_{\mu\mu} \cos \delta\vert_{0,\pi} \right]^2 \bigg\}
\label{eq:chi_dis}
\end{eqnarray}

In order to quantify the effects due to CP violation, another quantity 
called CP fraction $f(\sigma > 3)$ is often used. This  refers to the fraction 
of $\delta $ values for which CP violation can be determined above a  particular value of 
significance (here, $3\sigma$).  Being a fraction, $f (\sigma > 3)$ naturally lies between $0$ and $1$.

\begin{figure}[htb]
\centering
\includegraphics[width=\textwidth]
{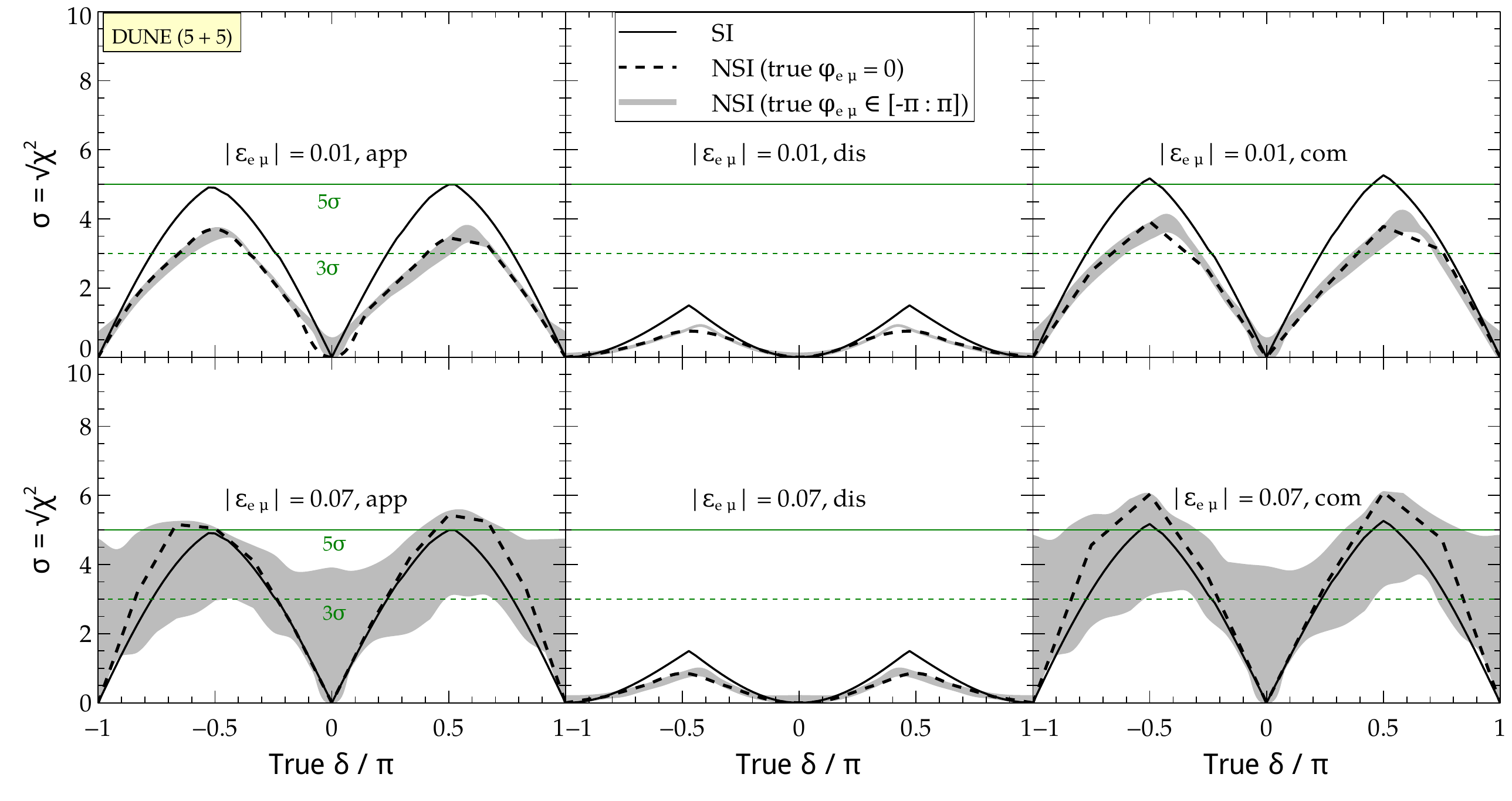}
\caption{\footnotesize{The impact of $\varepsilon_{e\mu}$ on the 
significance with which the CP violation can be determined as a function of the value of 
$\delta$ at \dune\, for an exposure of 350 kt.MW.yr assuming NH. The solid 
black curve represents the sensitivity for our reference design. 
Both the moduli and phases are varied as mentioned in the legend. The appearance and 
disappearance channels are shown separately and the sensitivity obtained by combining both the 
channels is also shown in the last column.
 }}
\label{fig:em}
\end{figure}

\begin{figure}[htb]
\centering
\includegraphics[width=\textwidth]
{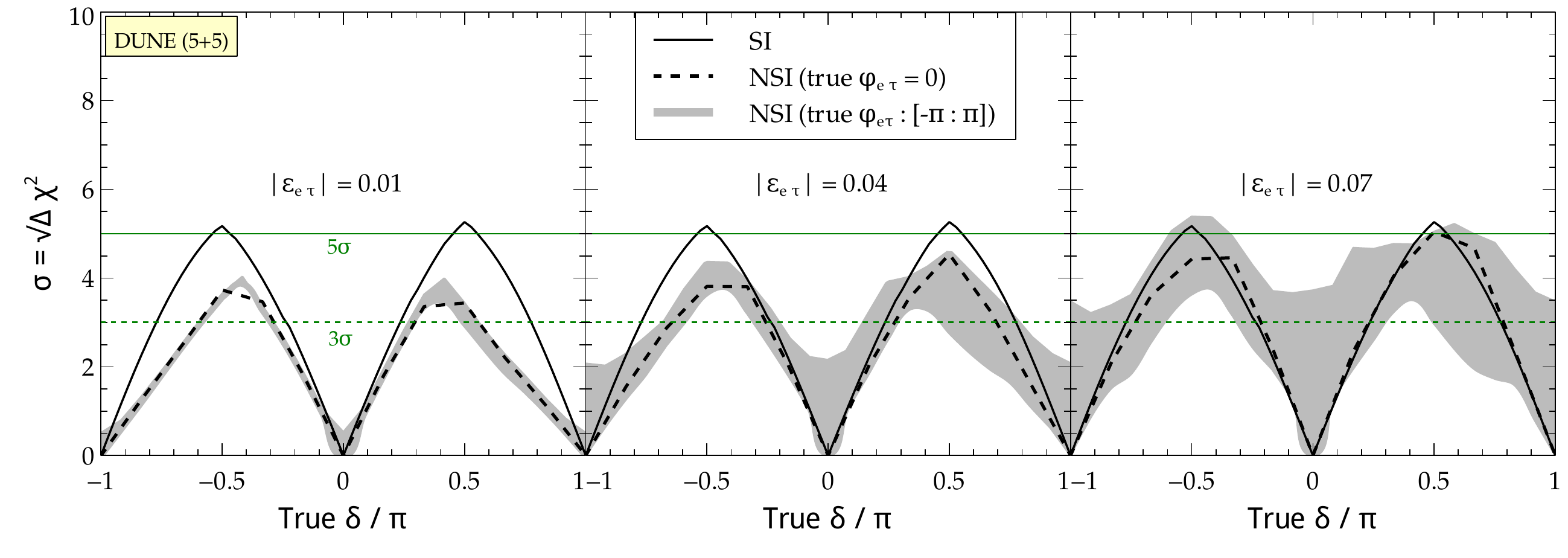}
\caption{\footnotesize{The impact of $\varepsilon_{e\tau}$ on the 
significance with which the CP violation can be determined as a function of the value of 
$\delta$ at \dune\, for an exposure of 350 kt.MW.yr assuming NH. The solid 
black curve represents the sensitivity for reference design. 
Both the moduli and phases are varied as mentioned in the legend. 
The combined sensitivity of appearance and disappearance channels 
is shown in the plot. 
 }}
\label{fig:et}
\end{figure}

\begin{figure}[hb!]
\centering
\includegraphics[width=\textwidth]
{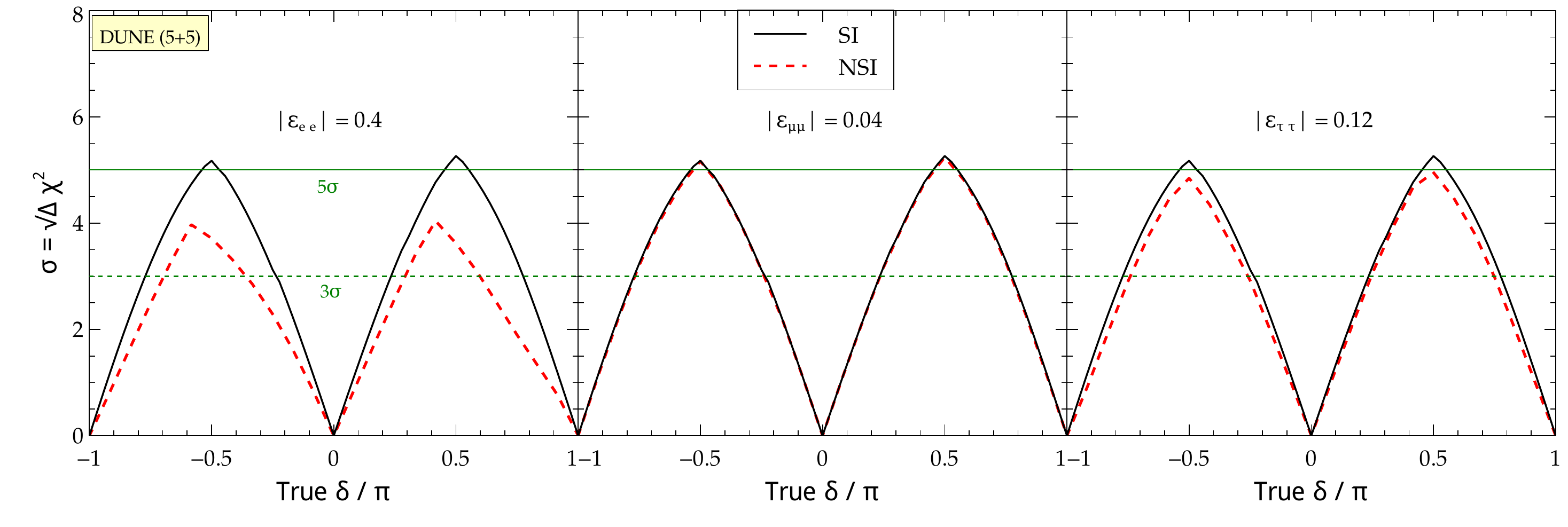}
\caption{\footnotesize{The impact of  $|\varepsilon_{e e}|$, 
$|\varepsilon_{\mu \mu}|$ and $|\varepsilon_{\tau \tau}|$ on the 
significance with which the CP violation can be determined as a function of true value of 
$\delta$ at \dune\, for an exposure of 350 kt.MW.yr assuming NH. The solid 
black curve represents the SI sensitivity for our reference design. 
The sensitivity obtained by combining  the appearance and disappearance 
channels is  shown.
 }}
\label{fig:ee}
\end{figure}

\section{Results}
\label{results}

\subsection{CP sensitivity  : impact of individual and collective NSI terms at DUNE}
\label{results_a}

In order to clearly understand the  impact of the NSI terms,  we first take only  one parameter non-zero at a time.  We show the effect of that particular parameter on CP sensitivity in the appearance  ($\nu_\mu \to \nu_e$)  as well as the  disappearance  ($\nu_\mu \to \nu_\mu$) channels. We also show the case with the two channels combined. 
  
Before we describe the impact of a particular NSI parameter (\ie~ $\varepsilon_{e\mu}$) we 
would like to point out that there are two effects responsible for 
  altering the value of the $\chi^2$ which compete with each other. 
  In general, we first note that NSI introduces more number of parameters (in the form of moduli of NSI terms 
  and the associated CP phases) in the sensitivity analysis and also introduces more sources of CP violation. One can have the following possibilities :
\begin{enumerate} 
\item[{\sl (a)}] Decrease in $\chi^2$ due to additional test values - If marginalization is carried out over more number of test parameters, it naturally results in a decreased value of $\chi^2$. This is purely a statistical effect. 
\item[\sl (b)] Increase in $\chi^2$ due to larger strength of true values -  In addition to more parameters in the test dataset (as mentioned in effect (a) above), one has to deal with a larger set of parameters in the true dataset as well. The variation over the values of the true NSI phases ($\varphi_{e\mu}$ or $\varphi_{e\tau}$) tends to broaden the grey band provided the true value of the moduli ($|\varepsilon_{e\mu}|$ or $|\varepsilon_{e\tau}|$) of the relevant NSI term is large.
 \end{enumerate}
In Fig.~\ref{fig:em},  we show the sensitivity to CP violation by exploiting appearance and disappearance   channels (in isolation and combined) for the off-diagonal NSI parameter,  
$\varepsilon_{e\mu}$ and compare it with the sensitivity obtained in case of SI as a benchmark. 
 The $3\sigma$ ($5\sigma$) value is shown as
 horizontal green dashed (solid) line to serve as a reference.
Let us first describe the SI case (shown as solid black curves). 
The  $P_{\mu e}$ channel dominates the sensitivity of CP violation which can be understood 
 from the presence of CP odd term.
   The mild CP sensitivity of the  $P_{\mu\mu}$ (due to the presence of CP even terms and absence of CP odd terms in $P_{\mu\mu}$) is
   not useful when considered in isolation but improves the $\chi^2$ in the combined case. The maximum (minimum) sensitivity in case of SI is 
   attained when $\delta \simeq \pm \pi/2$ ($\delta=0,\pm\pi$). In presence of NSI, in general the 
maximum $\chi^2$ is shifted from the SI maximum, $\delta = \pm \pi/2$. This is due to the shift in the   position of peaks and dips from the SI curve at the level of probability as mentioned in Sec.~\ref{framework_b}.

In $P_{\mu e}$, the presence of additional 
CP odd ($\sin \delta $ - like) terms  in presence of NSI  makes it possible for the 
 effect (b) to overtake  (a) if the value of the  NSI parameter is large enough.
If we see the top row of Fig.~\ref{fig:em}, the value of NSI parameter is small (true $|\varepsilon_{e\mu}| = 0.01$) and the black dashed curve
 (true $|\varepsilon_{e\mu}| \neq 0 $) 
and the grey band (true $|\varepsilon_{e\mu}| \neq 0 $, $\varphi_{e\mu} \in [\pi:\pi]$) are always below the SI case due to dominant (a) above. 
But for true $|\varepsilon_{e\mu}| = 0.07$, effect (b) becomes larger than effect (a) and we note that the NSI (with $|\varepsilon_{e\mu}| \neq 0$) overtakes SI. 
Also, the grey band spreads around the SI curve.  The most surprising outcome is that there can be $\geq 3\sigma$ sensitivity to CP violation 
  even when $\delta=0,\pm\pi$  (SI, CP conservation)  for some (un)favourable choice of NSI moduli and phases. 
  We can see this in the bottom panel of left and right plot of Fig.~\ref{fig:em}.  
  %
 As can be seen from the middle plots in the top and bottom rows of Fig.~\ref{fig:em} corresponding to $P_{\mu\mu}$, such an overtaking is not possible due to the absence of  
CP odd terms  (unless CPT is violated)  which forbids (b) to overtake  (a) and there is always a net reduction in $\chi^2$ due to (a) even if the NSI parameter is large.

In Fig.~\ref{fig:et}, the combined (appearance + disappearance) sensitivity to CP violation is shown both for SI and when the NSI parameter 
$\varepsilon_{e\tau}$ is incorporated. The effects are comparable in strength and 
 similar in nature to that of $\varepsilon_{e\mu}$ described above. The impact of the off-diagonal NSI parameter 
  $\varepsilon_{\mu\tau}$ on the CP sensitivity is found to be
 negligible even  if we choose values close to the upper limit mentioned in 
 Eq.~\ref{tinynsi} and hence is not shown here.

Having described the effect of off-diagonal NSI terms, we now address the impact of the diagonal ones - $\varepsilon_{ee},\varepsilon_{\mu\mu},\varepsilon_{\tau\tau}$.
We show the impact of  the three diagonal NSI parameters  ($\varepsilon_{ee}$, $\varepsilon_{\mu\mu}$ and $\varepsilon_{\tau\tau}$) in Fig.~\ref{fig:ee}. The effect of $\varepsilon_{\mu\mu}$ is very small as it is the most constrained parameter (Eq.~\ref{tinynsi}).
For the choice of values of the NSI parameters, the CP sensitivity sees a drop most likely
 due to the statistical effect (a) dominating in these cases. 

\begin{figure}[htb]
\centering
\includegraphics[width=\textwidth]
{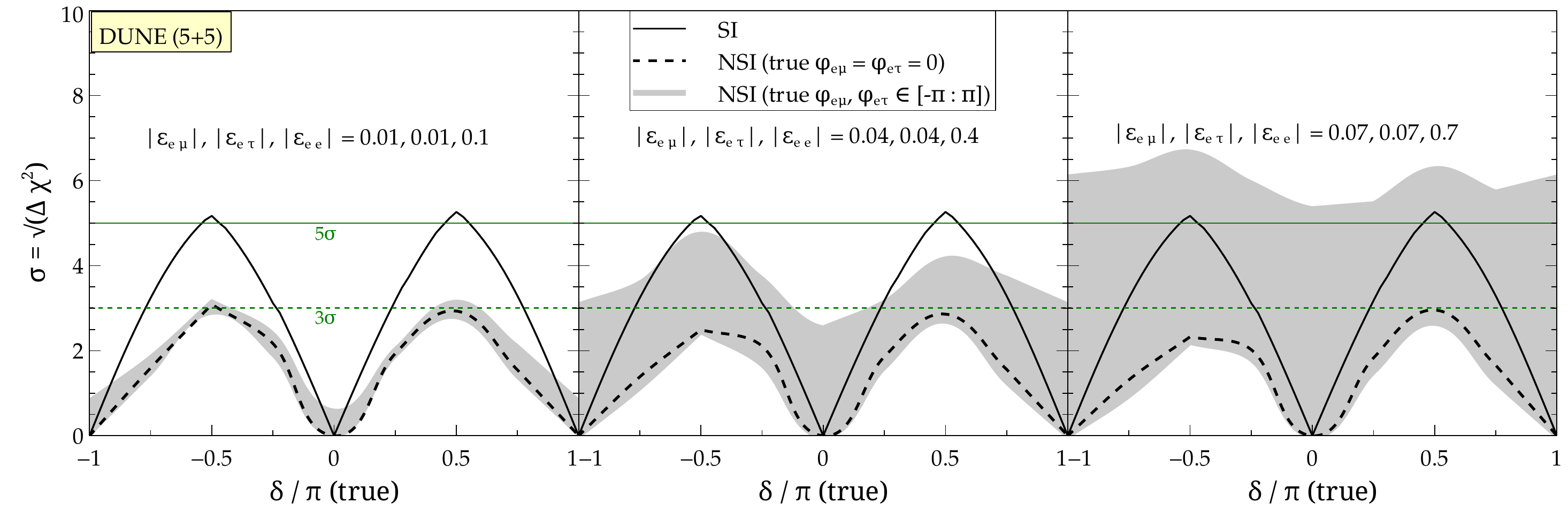}\caption{\footnotesize{CP sensitivity for collective 
NSI terms at \dune.
 }}
\label{fig:em_et}
\end{figure}

After understanding the impact of individual diagonal as well as off-diagonal NSI terms, we now address the collective effect of the most influential 
 NSI terms as far as  CP sensitivity is concerned. In Fig.~\ref{fig:em_et}, we show the collective impact of the three terms ($|\varepsilon_{e e}|$,  $|\varepsilon_{e \mu}|$,  $|\varepsilon_{e \tau}|$)
  which show the largest impact when considered in isolation. 
 We note that when the NSI terms are small, the associated phases of the NSI terms (even if taken collectively) do not contribute in an observable manner to (b) and (a) dominates. 
 However when we take somewhat larger values, we see the interplay of the two effects (a) and (b) with the possibility of second effect (b) overtaking the first (a) as we go from 
 small to large values keeping the marginalisation range intact.

 We summarize the impact of NSI on the CP violation sensitivity at long baselines as shown in Fig.~\ref{fig:em_et} for \dune.
  If we compare the solid and dashed black curves, we note  that for small values of parameters ($0.01,0.01,0.1$) NSI brings down the $\chi^2$ from $\sim 5 \sigma$ to $\sim 3\sigma$ at $\delta \sim \pm\pi/2$ for the case of zero NSI phases.   The impact of true non-zero  NSI phases can be seen in the form of grey bands for the choice of moduli of the NSI terms.  
  For larger values of parameters  ($0.07,0.07,0.7$)  NSI can drastically alter the $\chi^2$ not only at $\delta \simeq \pm\pi/2$ (SI, maximum) 
  but at almost all values of $\delta$ including at $\delta=0,\pm\pi$ if we allow for phase variation. For some particular choice of the NSI moduli and phases, we note that 
  in this case, the $\chi^2$ decreases from $\sim 5 \sigma$ to $\sim 2.5\sigma$ or increases to $\gsim 5.5 \sigma$ not only at $\delta \simeq \pm\pi/2$ but for most values of $\delta$. This can lead to a misleading inference that  CP is violated even when we have CP conservation in the SI case ($\delta=0,\pm\pi$).   
Here the phases have a bigger impact which can be seen as  widening of the grey bands as we go from smaller to larger moduli of NSI terms.

\subsection{Dependence on  $\theta_{23}$ and $\delta m^2_{31}$}
\label{results_b}

\begin{figure}[htb]
\centering
\includegraphics[width=\textwidth]
{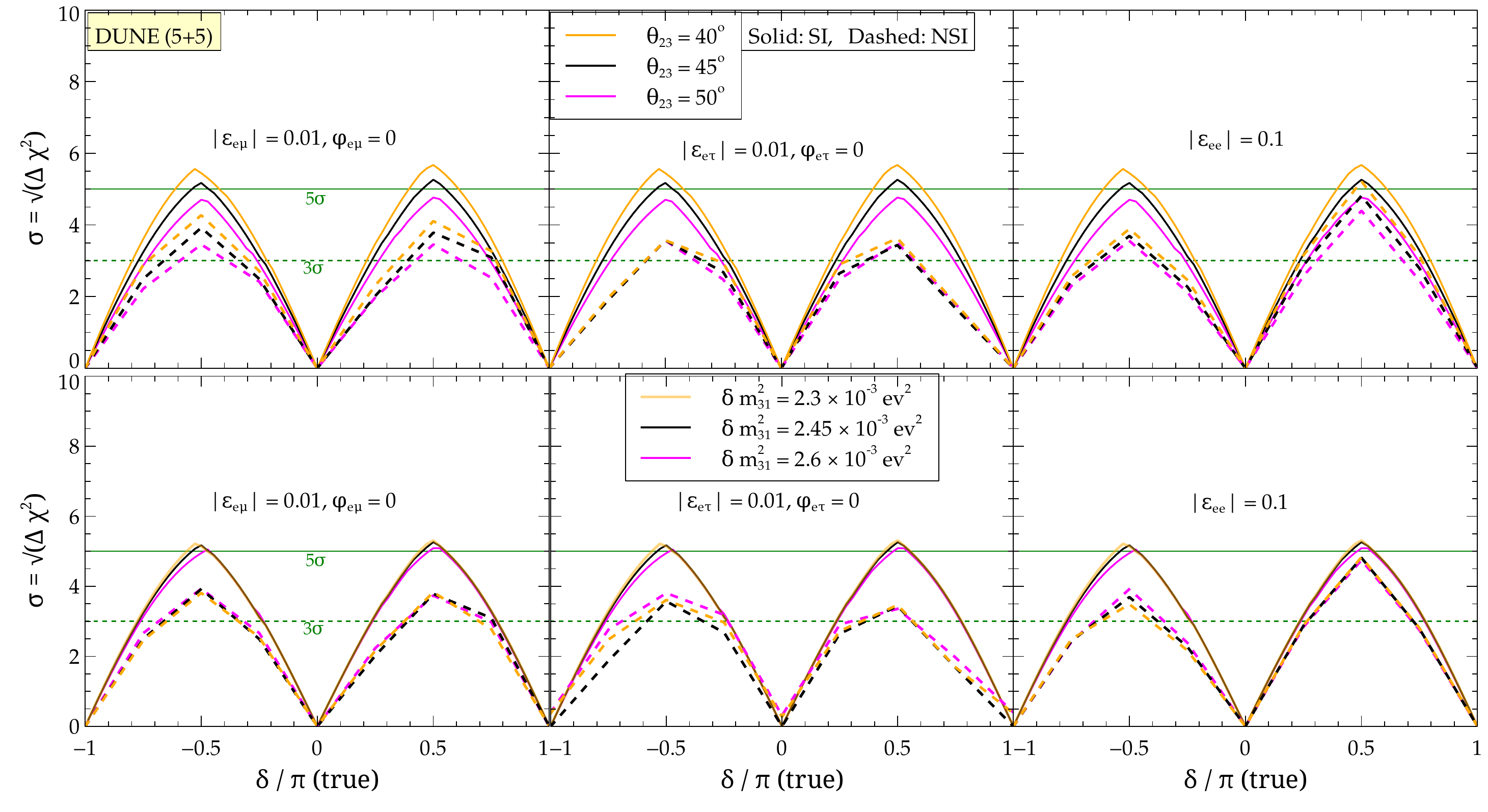}
\caption{\footnotesize{The dependence of CP sensitivity on the value of $\theta_{23}$ and $\delta m^2_{31}$ varied in the allowed range. The black curve is for 
$\theta_{23} = 45$ degrees and for our reference setup (provides a significance of at least $3\sigma$ for $\sim 55\%$ of $\delta$ values). 
 }}
\label{fig:th23d31}
\end{figure}

The variation in CP sensitivity due to different values of $\theta_{23}$ and 
$\delta m^2_{31}$  in the allowed range is shown in Fig.~\ref{fig:th23d31} for SI and NSI cases 
(zero NSI phases).  For $\theta_{23}$, as can be seen from the solid curves for SI, 
the significance (in presence of diagonal and off-diagonal NSI) decreases 
almost uniformly for all values of $\delta$ as $\theta_{23} $ becomes larger. This can be 
understood from Eqs.~\ref{pap}~\cite{Kimura:2002wd} and \ref{chisq}. The $P_{\mu e}$ 
increases with $\theta_{23}$ and therefore the $\chisq$ decreases. 
For no extra phases, we expect the sensitivity  in the presence of NSI  to be lower than the 
SI case due to the statistical effect. 
%

 For $\delta m^2_{31}$, the solid curves for SI show that
  the significance does not change significantly 
for all values of $\delta_{CP}$ as $\delta m^2_{31} $ is varied. Once again this can be 
understood from Eqs.~\ref{pap}~\cite{Kimura:2002wd} and \ref{chisq}. The true value of $\delta m^2_{31}$
does not impact  $P_{\mu e}$   and therefore the $\chisq$ remains almost the same. 

%

\subsection{Comparison with other experiments}
\label{results_c}

\begin{table}[htb]
\centering
\begin{tabular}{|l | c c | c c  |}
\hline
Experiment & \multicolumn{2}{c |}{Appearance channel} & \multicolumn{2}{c 
|}{Disappearance channel} \\
                                         \cline{2-5}
                                        & $\nu_\mu \to \nu_e$ 
                                        & 
                                         $\bar\nu_\mu \to \bar\nu_e$ & 
                                         $\nu_\mu \to \nu_\mu$ &
                                          $\bar \nu_\mu \to \bar\nu_\mu$                                                                                      
                                        \\ 
\hline
&&&&\\
\dune &&&&\\
$\delta=-\pi/2$ & 1610/1779 & 229/214 & 11431/11418 & 7841/7840 \\
 $\delta=0$ & 1350/1803 & 292/257 & 11401/11313 & 7805/7826 \\
$\delta=\pi/2$ & 1028/1182 & 309/279 & 11431/11417 & 7841/7840 \\
 &&&&\\
{{\nova}} &&&&\\
$\delta=-\pi/2$ & 90/95 & 17/15 & 142/143 & 46/47 \\
$\delta=0$ & 78/93 & 25/20 & 141/140 & 45/46 \\
$\delta=\pi/2$ & 57/62 & 30/27 & 142/143 & 46/47 \\
&&&&\\
{{\ttok}} 
&&&&\\
$\delta=-\pi/2$ & 127/130  & 20/19 & 372/371 & 130/129 \\
$\delta=0$ & 111/119 & 29/27 & 367/365 & 127/128 \\
$\delta=\pi/2$ & 80/82  & 33/32 & 372/371 & 130/129 \\
&&&&\\
{{\ttohk}} &&&&\\
$\delta=-\pi/2$ & 10231/10490 & 4882/4708 & 30048/30041 & 31321/31334 \\
 $\delta=0$ & 8979/9631 & 6989/6474 & 29641/29515 & 30920/31059 \\
$\delta=\pi/2$ & 6431/6643 & 7962/7737 & 30048/30041 & 31321/31334 \\
 \hline
\end{tabular}
\caption{\label{tab:events} Total number of signal events (SI/NSI) summed over all energy bins for each experiment using the oscillation parameters given in Table 1. 
For NSI, we show the collective case when the NSI parameters  $|\eem|=0.07, |\eet|=0.07, |\eee|=0.7$, $\varphi_{e\mu}=0$ and $\varphi_{e\tau}=0$  are considered.}
\end{table}


We now discuss how various currently running and future experiments that will aid  in determining the CP violation sensitivity in conjunction with \dune\, or in isolation. Before we go on, we 
give a brief description of the experiments that are sensitive to the appearance  ($\nu_\mu \to \nu_e$) channel as well as  the disappearance ($\nu_\mu \to \nu_\mu$) channel.  
\begin{description}
\item \ttok : The \ttok\, experiment has a baseline of $295$ km and the detector is placed at an off-axis (2.5 degrees) location.  
An intense beam of neutrinos  (mainly $\nu_\mu$ or $\bar\nu_\mu$) produced in the J-PARC accelerator facility in Tokai are directed 
towards the Super-Kamiokande detector (22.5 kton fiducial mass) situated in Kamioka. The near detectors located 280 m away from the point of neutrino production are  used to monitor  the neutrino flux. 
The $\nu_\mu$ beam peaks at $E\sim 0.6$ GeV which is close to the first oscillation maximum of $P_{\mu e}$.  The proton beam power is
 770 kW with proton energy of 50 GeV for 3 years (in $\nu$ mode) + 3 years (in $\bar \nu$ mode) which corresponds to  
a total exposure of $8.3 \times 10^{20}$ protons on target (p.o.t) per year.

\item \nova : The \nova\, experiment has a baseline of $810$ km and the detector is placed at an off-axis ($0.8$ degrees) location.
\nova\, stands for NuMI Off-axis $\nu_e$ Appearance experiment.  
An intense beam of neutrinos  (mainly $\nu_\mu$ or $\bar\nu_\mu$) produced by firing protons from FermiLab Main Injector on a graphite target.  
$\nu_\mu (\bar \nu_\mu)$ beamline is directed towards a  Totally Active Scintillator Detector (TASD) of fiducial mass 14 kton  placed in Ash River, 
Minnesota. 
This off-axis narrow-width beam peaks at $\sim 1.6$ GeV which is the energy at which $\nu_\mu \to \nu_e$ oscillaltion sees a maximum. 
A $0.3$ kton near detector is located at the FermiLab site to monitor the un-oscillated neutrino flux. The experiment will be running 
 in $\nu$ mode for 3 years and $\bar \nu$ mode for 3 years with a NuMI beam power of 0.7 MW and 
 120 GeV proton energy, corresponding to $6.0 \times 10^{20}$ p.o.t per year.

\item \dune : The \dune\, experiment has a baseline of $1300$ km and the detector is placed at an on-axis location.
A new, high intensity,  neutrino beam will be directed towards a  LArTPC located at Homestake at a distance of $1300$ km. 
The $\nu_\mu$ beam peaks at $E\sim 2.5$ GeV which is close to the first oscillation maximum of $P_{\mu e}$.   This
 facility is designed for   operation at a proton beam power of $1.0$ MW, with proton energy of $120$ GeV that will deliver $10^{21}$ p.o.t. in  $\sim 200$ days per calendar year.  
To have the LArTPC cross-sections, we have scaled the inclusive charged current  cross sections of water by $1.06 (0.94)$ for the $\nu (\bar \nu)$ case.

\item \ttohk : The \ttohk\, experiment has a baseline of $295$ km and the detector is placed at the same off-axis ($0.8$ degrees) location as in \ttok. 
The idea is to upgrade  the T2K experiment, with a much larger detector (560 kton fiducial mass) located in Kamioka so that much larger statistics is ensured. \ttohk\, will run for  1 year (in $\nu$ mode) + 3 years (in $\bar \nu$ mode).
The proton beam power is $7.5$ MW with proton energy of $30$ GeV that will deliver $1.6 \times 10^{22}$ p.o.t.  per  year. 
\end{description}
%

\begin{figure}[htb]
\centering
\includegraphics[width=.8\textwidth]
{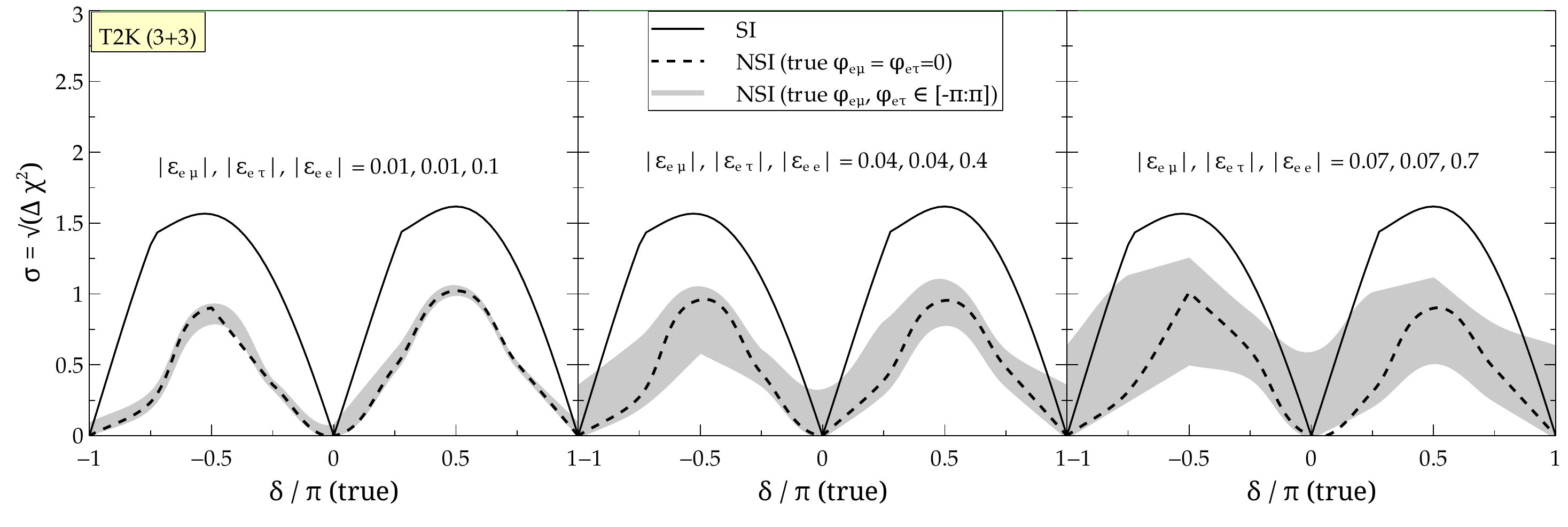}
\includegraphics[width=.8\textwidth]
{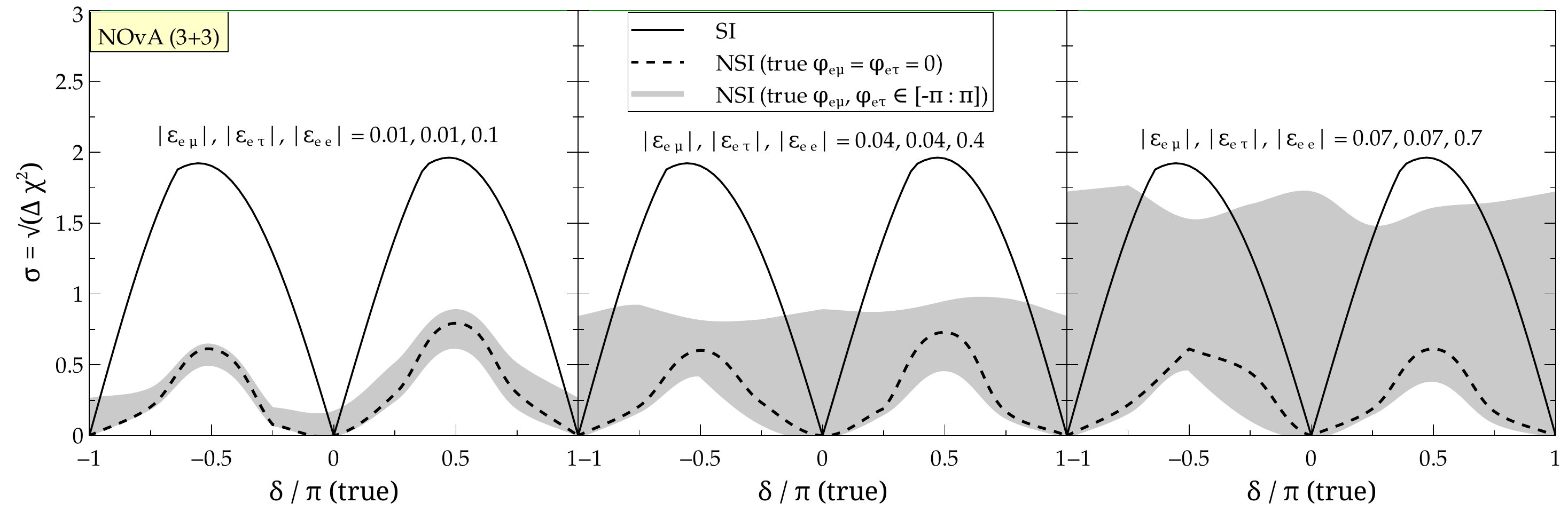}
\includegraphics[width=.8\textwidth]
{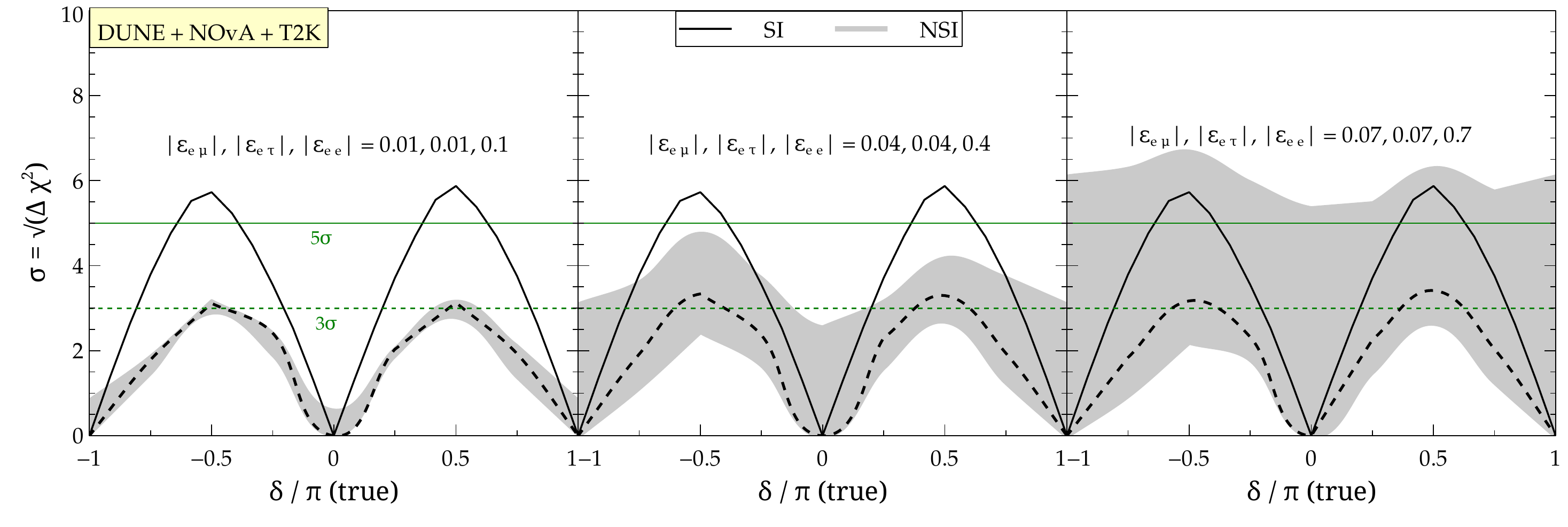}
\caption{\footnotesize{CP violation sensitivity at \ttok, \nova\, and \ttok+\nova+\dune\, 
 for collective NSI case and SI as a function of true $\delta$.
 }}
\label{fig:em_etn}
\end{figure}

\begin{figure}[htb]
\centering
\includegraphics[width=\textwidth]
{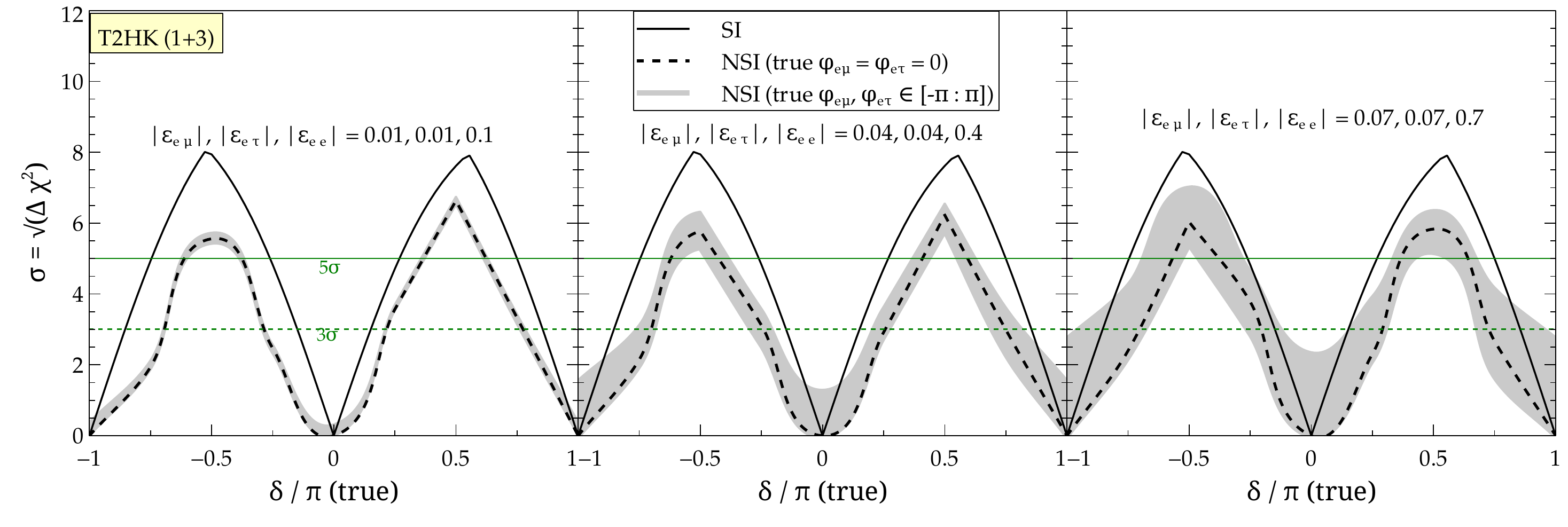}
\includegraphics[width=\textwidth]
{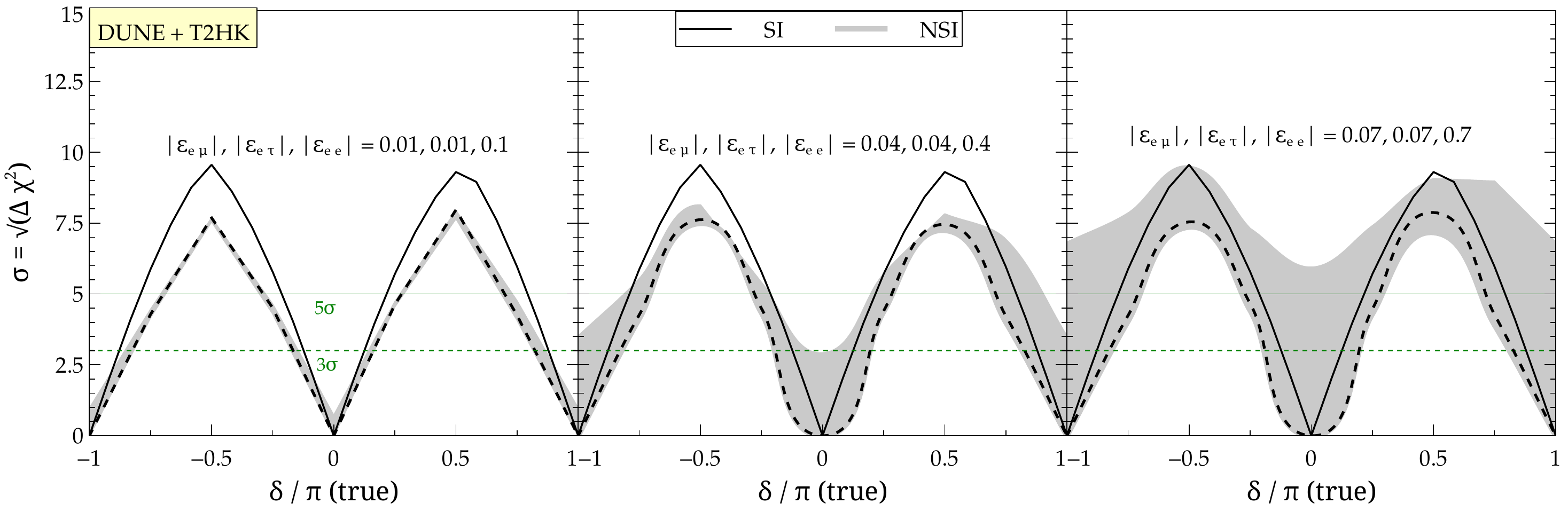}
\caption{\footnotesize{CP violation sensitivity  at \ttohk\, and \ttohk+\dune\, for collective 
 NSI case and SI as a function of true $\delta$.  
 }}
\label{fig:em_et_hk}
\end{figure}

The detailed detector characteristics and systematic errors are listed in Table~\ref{tab:sys}. In 
 Table~\ref{tab:events}, we list the energy integrated events~\footnote{Energy range for the various experiments
  is mentioned in Sec.~\ref{framework_c}} for the four experiments (using appearance and disappearance
  channels) for neutrinos as well as antineutrinos for NH. One striking feature to note is that the events
   in the disappearance channel are much larger than in the 
   appearance channel. This is due to the fact that the maximum value that $P_{\mu\mu}$ takes is close to 
   $1$ while $P_{\mu e}$ at best goes up to $\sim 0.1$. The larger detector size of \ttok\,  compensates for the 
    shorter baseline when compared to the smaller detector of \nova\,  with a longer baseline and the event rates for the two experiments are 
    comparable.  The event rates are some what larger for \dune\, as it has a bigger detector in comparison to
     \nova. But, \ttohk\, with its massive detector overcomes the limitation of baseline being short and
      gives the maximum number of events.

  The expected sensitivity offered by 
different experiments (singly or combined) is illustrated  in Figs.~\ref{fig:em_et}, \ref{fig:em_etn} 
and \ref{fig:em_et_hk}.  Fig.~\ref{fig:em_et} shows the CP violation sensitivity for \dune. 
 In Fig.~\ref{fig:em_etn}, we show the CP sensitivity for \ttok, \nova\, and a combination of \ttok,  \nova\, and \dune. Finally, we 
 show the CP violation sensitivity for \ttohk\, in Fig.~\ref{fig:em_et_hk} which is competitive with \dune.

We have shown the CP violation sensitivity at \dune\,  in Fig.~\ref{fig:em_et} and discussed the features   
in Sec.~\ref{results_a}. In  Fig.~\ref{fig:em_etn}, we show the CP sensitivity for  \ttok (top row), \nova (middle row) as well as
 a combination of \ttok\,, \nova\, and \dune (bottom row).
As in Fig.~\ref{fig:em_et}, the characteristic double peak is seen for all the three cases in Fig.~\ref{fig:em_etn}.
 If we now look at \ttok\, and \nova\,  individually, we note 
 that the CP violation sensitivity almost never reaches $3\sigma$ (it barely touches $\sim 1.6\sigma$ (for \ttok) and $\sim 1.8\sigma$ (for \nova)). 
  This means that these two current experiments considered in isolation are not so much interesting as far as 
  CP violation sensitivity is concerned. 
  This does not come as a surprise as these are not optimized for CP sensitivity. 
  However, if we combine data from these two experiments with \dune, we note 
  that CP violation sensitivity improves slightly  (from $\sim 5.1\sigma$ to $\sim 5.6\sigma$ in the SI case near the peak. For NSI  (zero NSI phases, dashed black curve) it improves marginally 
   from $\sim 3\sigma$ to $\gsim 3\sigma$. In general, we note that the if phases are taken into account, the grey bands expand and even 
   out as we go from small to large NSI, the peaks at $\delta \sim \pm \pi/2$ smoothen out 
    which means that there is no clear demarkation of CP conserving ($\delta = 0,\pm\pi$) and CP violating  values of $\delta$.

\begin{figure}[htb]
\centering
\includegraphics[width=\textwidth]
{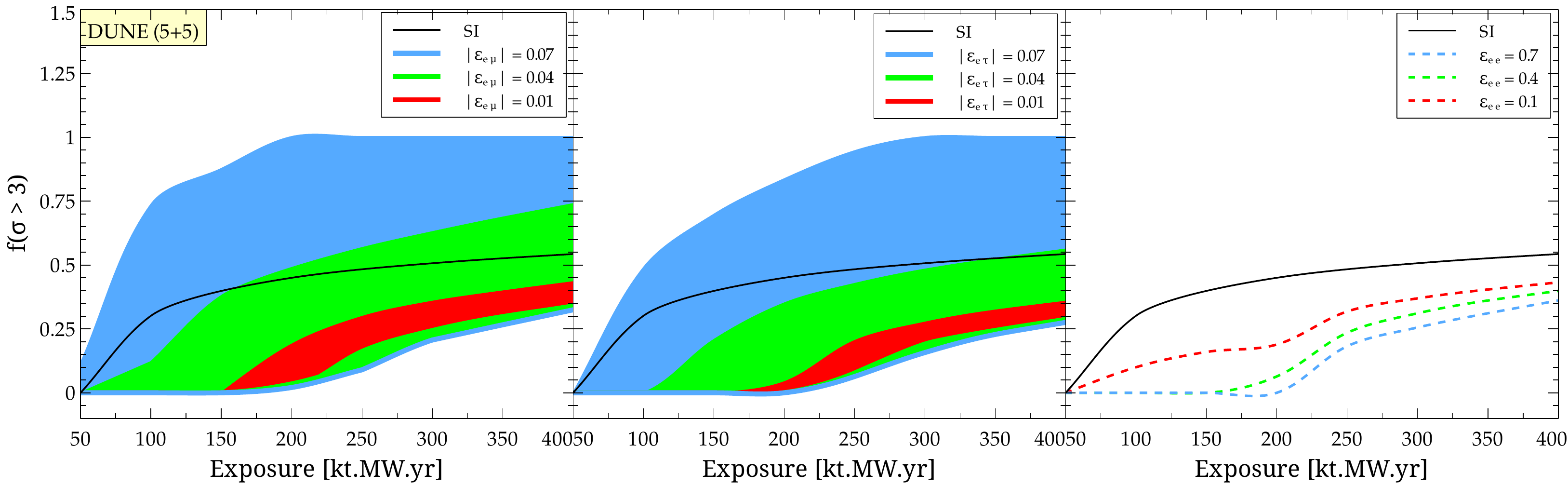}
\caption{\footnotesize{The 
CP fraction $f (\sigma > 3)$  for which the 
sensitivity to CP violation is 
greater than $3 \sigma$  as a function of exposure for SI and NSI case assuming NH. The three plots 
correspond to three different NSI parameters taken one at a time with full phase variation. The red, green and blue shaded regions correspond to different values of $\eem$ and $\eet$. 
 }}
\label{fig:single_nsi_3sigma}
\end{figure}
 
 In Fig.~\ref{fig:em_et_hk}, we show the CP violation sensitivity for \ttohk. We note that \ttohk\, offers CP sensitivity that is competitive with \dune\, individually as well as \ttok, \nova\, and \dune\, combined (SI and NSI both).  
 This can be ascribed to the high statistics offered by the HK.  Near the peak, we note that it can go upto $\sim 8\sigma$ for SI and $\gsim 5 \sigma$ for NSI (zero phases). 
 Another intriguing feature from \ttohk\, panel is that the NSI phases do not have as dramatic effect as seen for \dune\,  when the  NSI terms are large  - this can be seen as shrinking of the grey regions in Fig.~\ref{fig:em_et_hk} (top panel, right most plot). This is due to the fact that the baseline of $295$ km is way too short for matter effects (SI and NSI both) to develop and play a significant role\footnote{Similar feature can also be seen from the \ttok\, panel in Fig.~\ref{fig:em_etn}.}.{{ This demonstrates the complementarity of bigger detectors (\ttohk) vis-a-vis the long baselines involved (\dune) where no clear demarkation of CP conserving ($\delta = 0,\pm\pi$) and CP violating values of $\delta$ was noticed. }}

\subsection{Optimal exposure for CP violation discovery}
\label{results_d}

The previous set of plots were obtained by keeping the total exposure fixed  for a given experimental configuration. 
 The maximum value of $f (\sigma >  3)$ guides the choice of optimal exposure for CP violation discovery.
Let us see how the choice of optimal exposure in case of SI is arrived at. We  have  already noted that the CP violation sensitivity as a function of $\delta$ has a double peak structure for SI due to the vanishing of the sensitivity 
at CP conserving values of $\delta$ ($=0,\pm\pi$). Therefore it is expected that none of the experiments considered in the present work can lead to a $100\%$ coverage in $\delta$ in the SI case. This no longer holds in presence of NSI.

 In Fig.~\ref{fig:single_nsi_3sigma}, we show the 
CP fraction for which the 
sensitivity to CP violation exceeds $3 \sigma$ as a function of exposure, labelled as $f (\sigma > 3)$. 
Let us first understand the SI case, we note that 
$f(\sigma > 3)$ rises  from $0 $ to $ \sim 0.4$ as a function of exposure  initially  
as we go from $50 - 150$ kt.MW.yr but saturates to a value $f(\sigma > 3) \simeq 0.5-0.55$ 
as we go to exposures beyond $\sim 350$ kt.MW.yr. Increasing the exposure further does 
not change this value drastically beyond $f(\sigma > 3) \simeq 0.5$. This is not unexpected as we have already noticed that it    is challenging to exclude those values of CP phase which lie
    close to the CP conserving values (\ie, $0$ and $\pi$). 
    So, in case of SI, the choice of optimal exposure is expected to be  
 $\simeq 350$ kt.MW.yr.



Let us now discuss  the impact of NSI on the choice of the optimal exposure. 
\begin{table}[htb]
\centering
\begin{tabular}{|l | c c | c c| }
\hline
NSI term & \multicolumn{2}{c |}{NH} & \multicolumn{2}{c |}{IH} \\
                                         \cline{2-5}
                                        & $f(\sigma >3 )$ (NSI)
                                        & $f(\sigma >3 )$ (SI)  & 
                                                                 $f(\sigma >3 )$ (NSI) &                          
                                              $f(\sigma >3 )$ (SI)                                   
                                        \\ 
\hline
%
 $|\eem|=0.01$ & $0.32 - 0.40$ & 0.52 & $0.35 -0.42$ & 0.58 \\
 $|\eem|=0.04$ & $0.30 - 0.69$ & 0.52 & $0.33 - 0.78 $ & 0.58 \\
 $|\eem|=0.07$ & $0.27 - 1.00$ & 0.52 &  $0.32 - 1.00 $ & 0.58 \\
 \hline               
 $|\eet|=0.01$ & $0.26 - 0.32$ & 0.52 & $0.23 - 0.32 $ & 0.58 \\
 $|\eet|=0.04$ & $0.24 - 0.53$ & 0.52  & $0.22-0.84$ & 0.58\\
 $|\eet|=0.07$ & $0.23 - 1.00$ & 0.52 & $0.21-1.00 $ & 0.58 \\
  \hline                 
 $\eee=0.01$ & $0.40$ & 0.52 & $0.36$ & 0.58 \\
 $\eee=0.04$ & $0.36$ & 0.52 & $0.30$ & 0.58 \\
 $\eee=0.07$ & $0.31$ & 0.52 & $0.27$ & 0.58 \\
 \hline
\end{tabular}
\caption{\label{tab:exposure}  $f (\sigma >3)$ at an exposure of $350$ kt.MW.yr for \dune\, using 
nominal systematics (see Fig.~\ref{fig:single_nsi_3sigma}).}
\end{table}
   For the NSI case, the three panels  in Fig.~\ref{fig:single_nsi_3sigma} correspond to the three different NSI terms (taken in isolation). 
  There are three coloured regions (blue, green, red) 
 for the off-diagonal NSI terms which correspond to the three values of moduli of 
 NSI parameters along with their respective phase variation 
 (analogous to the grey bands seen in Fig.~\ref{fig:em} and \ref{fig:et}). 
 For the diagonal NSI terms, there are three dashed
 lines (blue, green, red) corresponding to three different values of diagonal 
 NSI parameter $\eee$ (see Fig.~\ref{fig:ee}). 
The plot on the left shows the impact of $\eem$. Even with the phase variation, 
 $f (\sigma > 3)$ (shown as red band) remains below the SI curve for small value of $\eem$ ($|\eem| = 0.01$). 
This is  due to  the dominating 
 statistical effect (a) mentioned in Sec.~\ref{results_a}.  
 $f (\sigma > 3)$ stays at zero and does not rise until an exposure of $\simeq 
 150$ kt.MW.yr is reached. Finally, it attains a value in the range 
 $\simeq 0.35-0.45$ as we reach exposures  
 $\sim 400$ kt.MW.yr. 
For intermediate and large values of $\eem$ ($|\eem|=0.04,0.07$) on the other hand, 
 $f (\sigma > 3)$ gets distributed over a larger range of values for some 
favourable choice of parameters (due to the interplay of (a) and (b) mentioned 
in Sec.~\ref{results_a}) as can be seen from the green and blue bands. 
 Incorporating the phase variation of the NSI parameter leads to an increase in the 
  value of $f (\sigma > 3)$ and it can reach $\sim 1$ when the exposure is barely 200 kt.MW.yr 
  (for some choice of phases, some part of the grey band is above the $3 \sigma$ line
   in Fig.~\ref{fig:em} and \ref{fig:et} for all true values 
  of $\delta$).  
  Similar effects are seen for the other off-diagonal parameter, $\eet$ which is shown in the middle panel. 
However for the diagonal NSI parameter $\eee$ (which is real), we note that 
the $f (\sigma > 3)$  (blue, green and red dashed lines) is always 
smaller than in the SI case for a given choice of systematics (see also Fig.~\ref{fig:ee}). 
This is again due to the statistical effect. 

  We have checked that if we take the true hierarchy as inverted hierarchy (IH) instead of NH, 
the impact of NSI shown in Fig.~\ref{fig:single_nsi_3sigma} is grossly the same (see Table~\ref{tab:exposure}). 
The impact of individual NSI terms on the value of $f (\sigma > 3)$ at an exposure 
of $350$ kt.MW.yr (which is the optimal choice for SI) at \dune\,  is  listed in Table~\ref{tab:exposure} for NH and IH. 
  
\subsection{Role of systematics}
\label{results_e}
  
\begin{figure}[htb]
\centering
\includegraphics[width=\textwidth]
{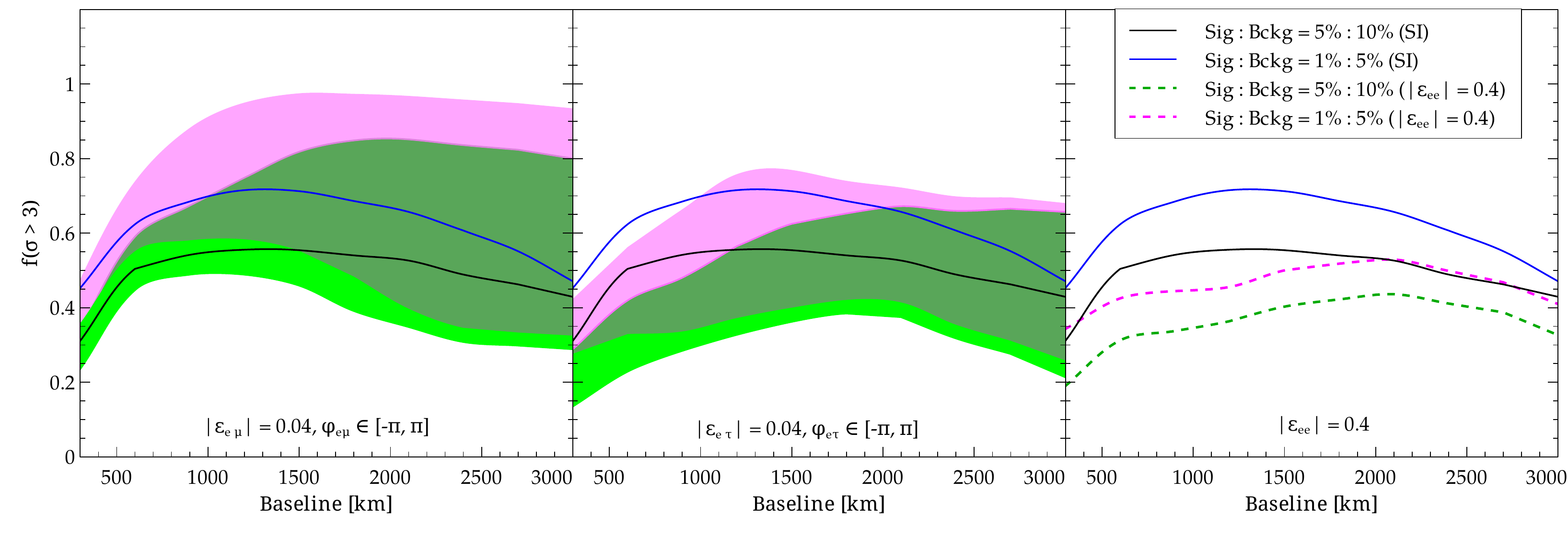}
\caption{\footnotesize{The 
CP fraction for which the sensitivity to CP violation is greater than $3 \sigma$ as a function of baseline for SI and NSI case. 
The black and blue solid curves correspond to the different systematics assumed for SI. 
The three plots correspond to three NSI parameters taken one at a time. 
The green (magenta) band corresponds to the choice of nominal 
(optimal) systematics with full phase variation for the off-diagonal NSI parameters while the green (magenta) dashed line corresponds to $\eee$ for nominal (optimal) systematics.
 }}
\label{fig:baseline_optimization}
\end{figure}
  %
%
The impact of different assumptions on systematics can be seen in 
  Fig.~\ref{fig:baseline_optimization}. 
The nominal set of systematics  is mentioned in Table~\ref{tab:sys}.
The black solid curve represents our nominal choice of systematics given 
in Table~\ref{tab:sys} while the blue solid curve is for an optimal choice mentioned in the legend~\cite{Bass:2013vcg}. The green (magenta) band 
corresponds to NSI case for off-diagonal parameters 
$\eem,\eet$ with full phase variation  for nominal (optimal) choice of systematics. The green (magenta) dashed curve is for $\eee$ for nominal (optimal) choice of systematics.

It can be seen that $f (\sigma > 3)$ nearly reaches its maximum  ($\sim 0.55$) possible value at around $1300$ km for SI (see Fig.~\ref{fig:baseline_optimization}). This implies that for the given 
configuration of the far detector planned for \dune\, (see Table~\ref{tab:sys}), 
the optimal distance to be able to infer the highest fraction of the 
 values of the CP phase is $\sim 1300$ km. 
Clearly,  even in case of SI, better systematics is expected to lead to a 
 larger $f (\sigma > 3)$ for a given baseline, say at 1300 km - 
 it changes from $\sim 0.55$ to $\sim 0.71$. For the SI case, better systematics ensures better detectability of 
 CP violation quantified in terms of fraction $f (\sigma > 3)$ and at the same time, 
 does not alter the optimal baseline choice for CP violation sensitivity. 
 In case of NSI, the green (magenta) band show the effect of two choices of systematics and there is an overlap between them as well as with the SI values. 
{\sl{These aspects  play a crucial role in altering the choice of best baseline for CP violation sensitivity.}} 
 However, in presence of NSI,  for the choice of NSI phases representing the  top (bottom) edge of the green or magenta band (we have used the dashed green or magenta lines to depict the diagonal NSI terms), the optimal choice of baseline ($L_{opt}$) that maximizes the CP fraction changes as a function of systematics (see Table~\ref{tab:system}).
%


\begin{table}[htb]
\centering
\begin{tabular}{|l | r r | r r |}
\hline
NSI term & \multicolumn{2}{c |}{Nominal systematics (green)} & 
\multicolumn{2}{c|}{Optimal systematics (magenta)} \\
                                         \cline{2-5}
                                        & NSI
                                        & SI  & 
                                           NSI &                          
                                             SI                                  
                                        \\ 
                                         \cline{2-5}
                                        & $f(\sigma >3 )$ $L_{opt}$
                                        & $f(\sigma >3 )$ $L_{opt}$  & 
                                                                 $f(\sigma >3 )$ 
                                                                 $L_{opt}$ &                          
                                              $f(\sigma >3 )$ $L_{opt}$                                  
                                        \\ 
& km & km & km & km\\
%
%
\hline                 
 $|\eem|=0.04$ & $0.85$ ($1800-2500$)   & 0.52 ($1300$) & $0.97$ ($1500-3000$) & 0.71 (
 $1300$) \\
 & $0.49$ ($800-1300$)  & & $0.59 $ ($800-1300$) &\\
 \hline                 
  $|\eet|=0.04$ & $0.65$ ($2000-3000$)   & 0.52 ($1300$) & $0.77$ ($1300-1500$) 
  & 0.71 ($1300$) \\
 & $0.37$ ($1800-2000$)  & & $0.40$ ($1800-2000$) &\\
  \hline                  
 $\eee=0.04$ & $0.43$ ($1900-2100$) & $0.52$ ($1300$) & 
 $0.52$ ($1900-2100$ ) & $0.71$ ($1300$)\\
 \hline
\end{tabular}
\caption{\label{tab:system} Maximum $f (\sigma >3)$ and optimal baseline range ($L_{opt}$) for the two different choices of 
 systematics (see Fig.~\ref{fig:baseline_optimization}) for NH. The values with larger (smaller) $f(\sigma > 3)$ correspond to upper (lower) edge of the respective bands. }
\end{table}


\subsection{CP violation sensitivity assuming known source}
\label{results_f}

\begin{figure}[htb]
\centering
\includegraphics[width=0.8\textwidth]
{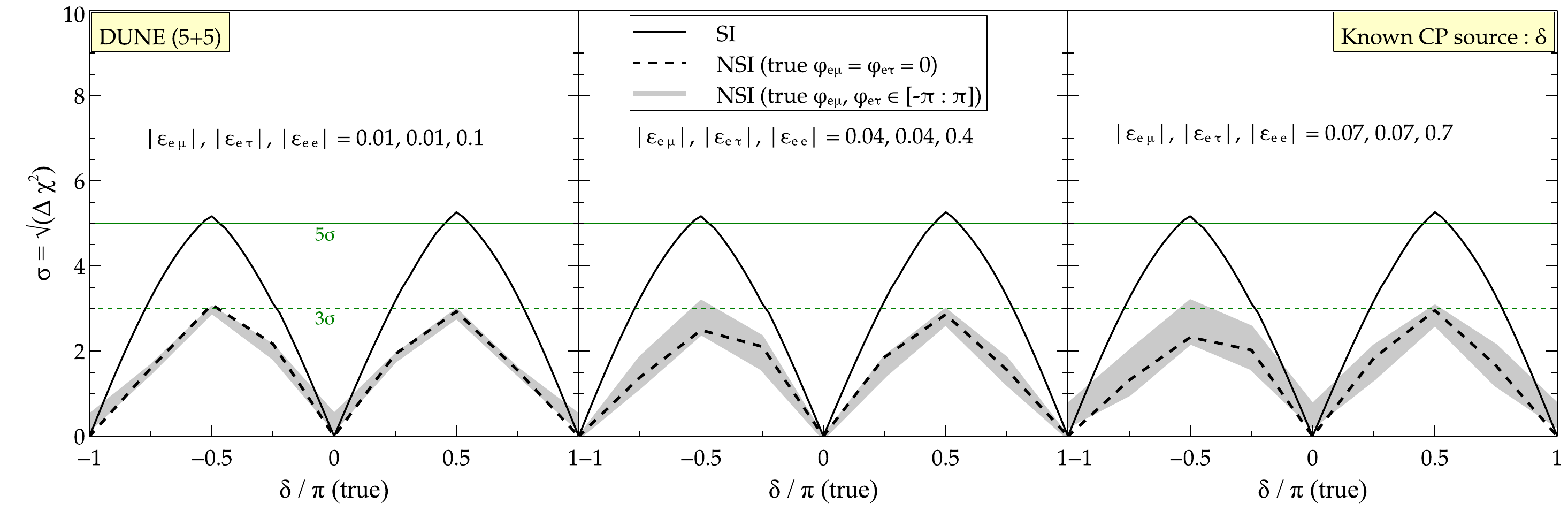}
\includegraphics[width=0.8\textwidth]
{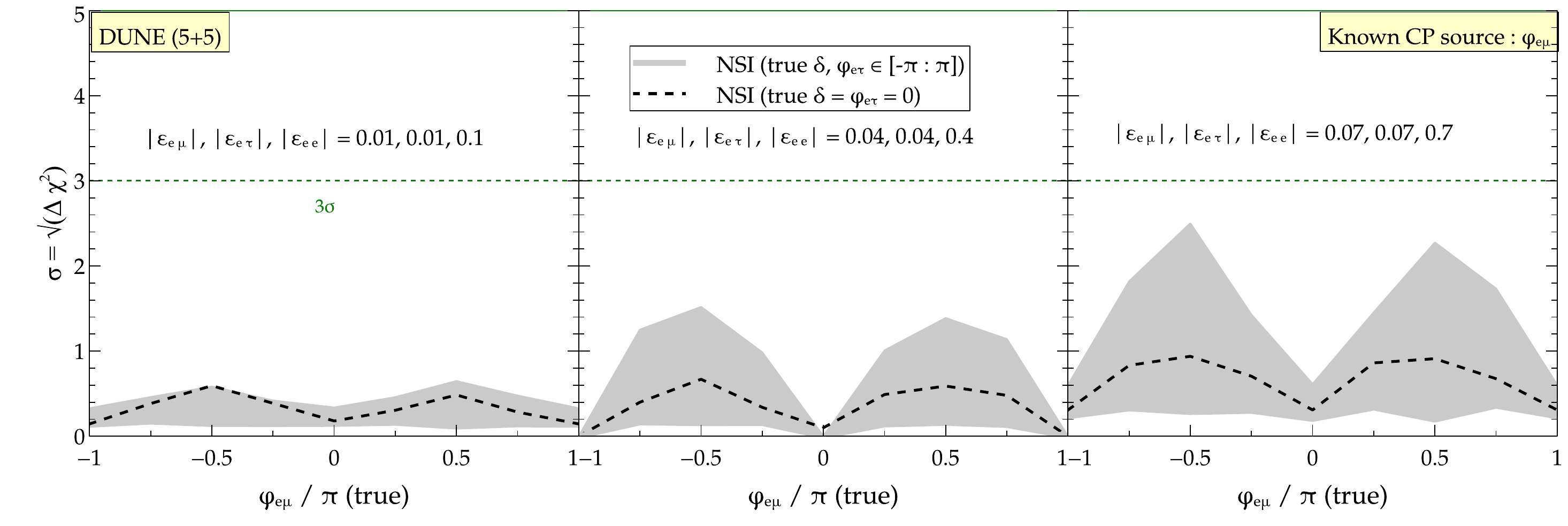}
\includegraphics[width=0.8\textwidth]
{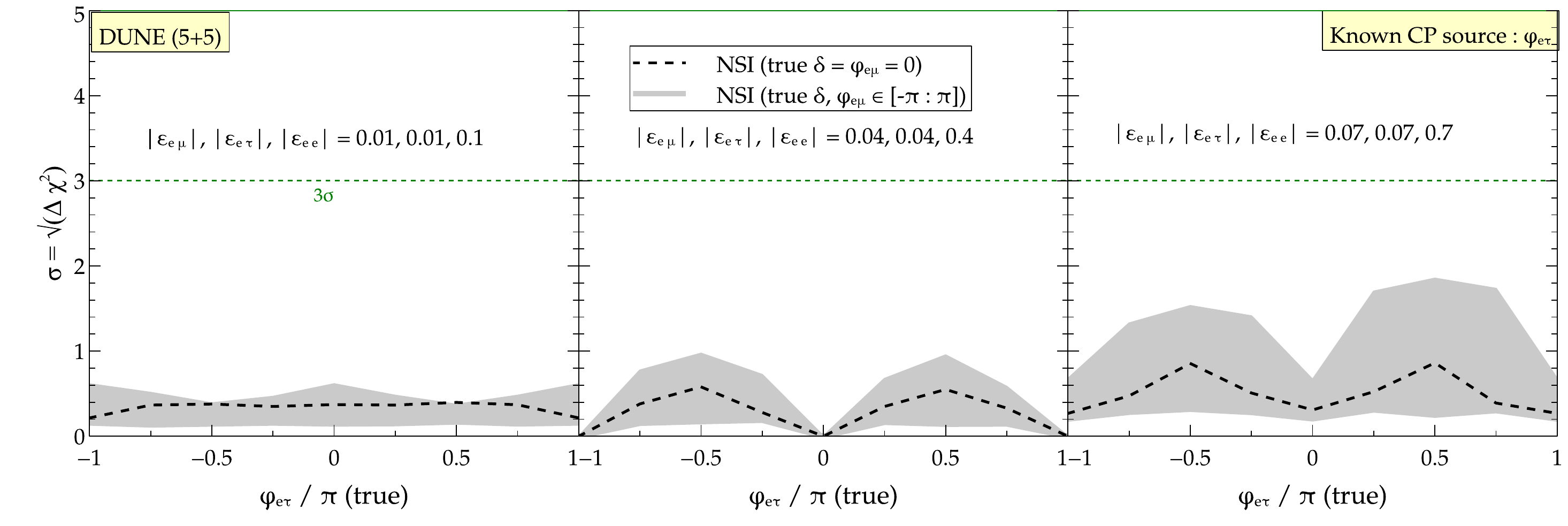}
\caption{\footnotesize{CP violation sensitivity if the source of CP violation is known. The three rows correspond to one known source (shown along the $x$-axis) at a time. 
 }}
\label{fig:fig8}
\end{figure}

In the preceeding discussion,  we assumed that the source of CP violation (i.e. whether it arises due to SI CP violating parameter $\delta$ or due to NSI CP violating parameters $\varphi_{e\mu}$, $\varphi_{e\tau}$) cannot be traced or in other words is unknown. If one knows which parameter is responsible for CP violation, the results are drastically modified. In order to illustrate this, let us assume that out of the possible sources of CP violation i.e. $\delta$, $\varphi_{e\mu}$ or $\varphi_{e\tau}$, only one of these is responsible at a time for generating CP violating effects. The results for the case of known source of CP violation are depicted in Fig.~\ref{fig:fig8}. The three rows correspond to the case when the known source is $\delta$ (top), $\varphi_{e\mu}$ (middle) and $\varphi_{e\tau}$ (bottom). 
%
From the top row, we note that the presence of NSI spoils the CP violation sensitivity irrespective of the strength of the NSI terms as compared to the standard case (black solid curves). Also, since $\delta$ is the only source of CP violation, in the first row, the CP violation sensitivity drops nearly to zero at $\delta =0,\pm\pi$ for all the three cases shown. This is in sharp contrast with Fig.~6 where all of the phases would have contributed to the CP violation sensitivity. The middle and the bottom row correspond to the case when $\varphi_{e\mu}$ and $\varphi_{e\tau}$ are the only sources of CP violation. Again the characteristic double peak structure is visible but the CP violation sensitivity is not as large as for top row simply because NSI effects are subdominant.
For Fig.~\ref{fig:fig8}, in the $\chi^2$ evaluation, we marginalize over the full allowed range~\footnote{The full allowed range is $[-\pi,\pi]$ in contrast to the CP conserving values $(0,\pi)$ considered in the previous section when the source of CP violation was assumed to be unknown. The statistical effect is expected to dominate in this scenario due to the wider ranges considered for the marginalisation over the test values of the CP violating parameters.} of the remaining phases (for the case of top row, $\varphi_{e\mu}$, $\varphi_{e\tau}$) in Eq.~\ref{chisq}.

%

\subsection{Measuring the phases that may be responsible for CP violation}
\label{results_g}
\begin{figure}[htb]
\centering
\includegraphics[width=0.4\textwidth]
{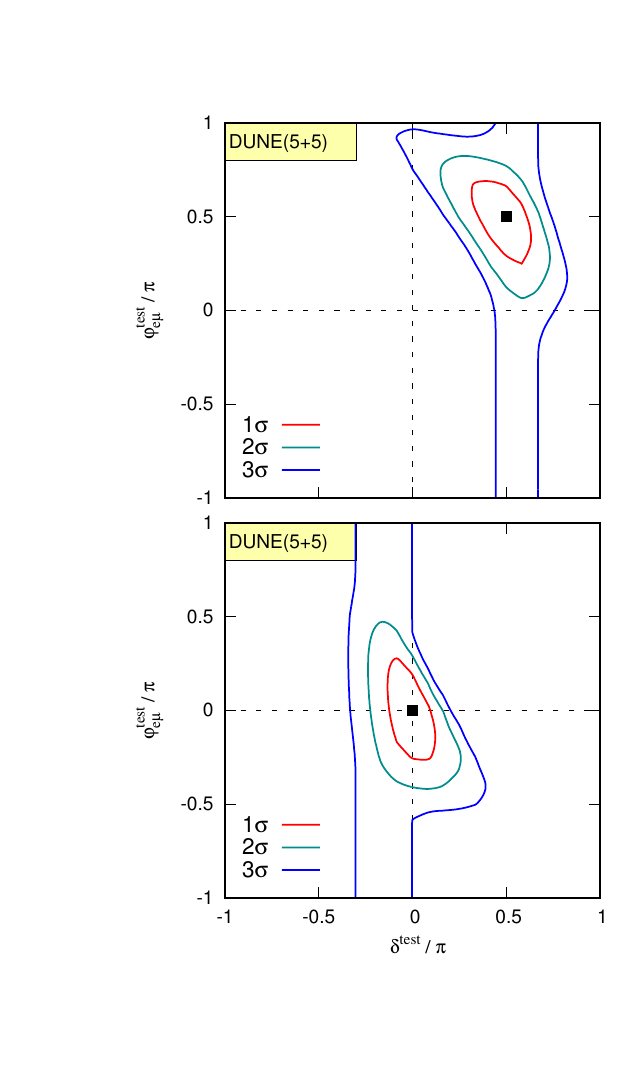}
\includegraphics[width=0.4\textwidth]
{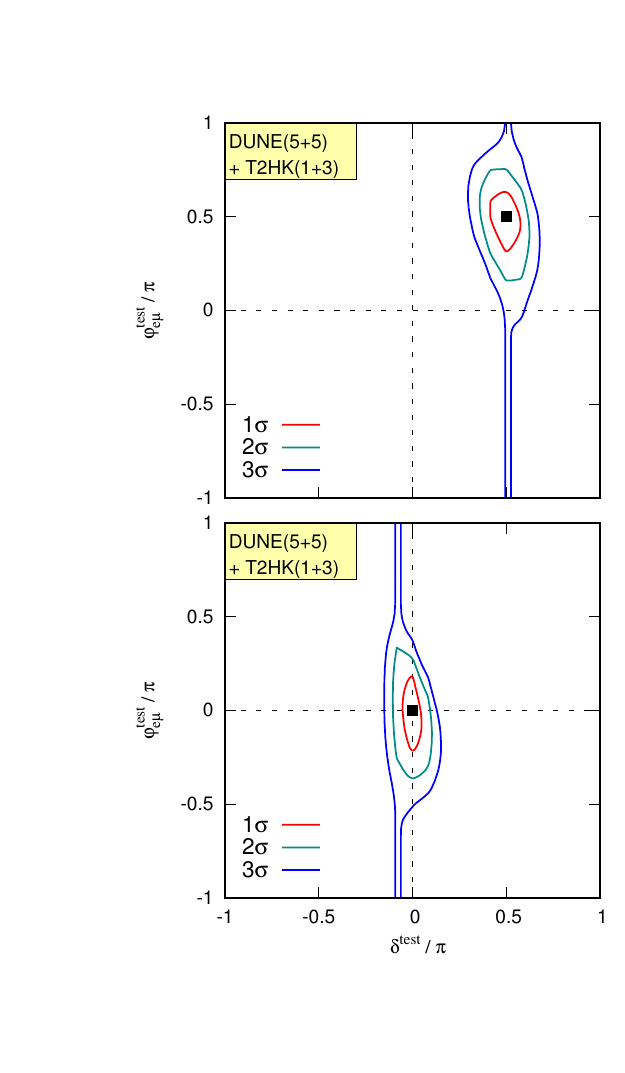}
\caption{\footnotesize{Regions in $\varphi^{test}_{e\mu} - \delta^{test}$ plane. 
The black dot represents the pair of true values $\{\varphi_{e\mu}^{true},
\delta^{true}\}$ which are taken to be $\{\pi/2,\pi/2\}$ (CP violating) in top row or
 $\{0,0\}$ (CP conserving) in the bottom row. The value of NSI parameter is taken to be 
 $|\varepsilon_{e\mu}|=0.04$. The plots on the left are for
  \dune\, and those on the right are for \dune\, + \ttohk. 
 }}
\label{fig:fig13}
\end{figure}

Independent of the question of the CP violation sensitivity that we have addressed in the present article, one can ask if it is possible to measure the CP phases at  long baseline experiments.
For the sake of simplicity, we assume that only one NSI parameter contributes at a time (let us assume that this is given by $\varphi_{e\mu}$)\footnote{For the other NSI parameter $\varphi_{e\tau}$, the results are similar.}. Let us now take some representative values of the true CP phases and discuss how well we are able to reconstruct those values among the allowed test ranges. In Fig.~\ref{fig:fig13}, for two possible choices of the pair of phases $\{\delta^{true},\varphi_{e\mu}^{true}\} = \{\pi/2,\pi/2\}$ (maximal CP violation) and $\{\delta^{true},\varphi_{e\mu}^{true}\} = \{0,0\}$ (CP conservation), we show the ability of \dune\, to reconstruct those phases assuming NH. 
%
 %
 For a comparison, we also show the results for the combined case of \dune\,+\ttohk\, where we see that the regions enclosed by the contours become narrower. 
 
 %


\begin{figure}[htb]
\centering
\includegraphics[width=0.6\textwidth]
{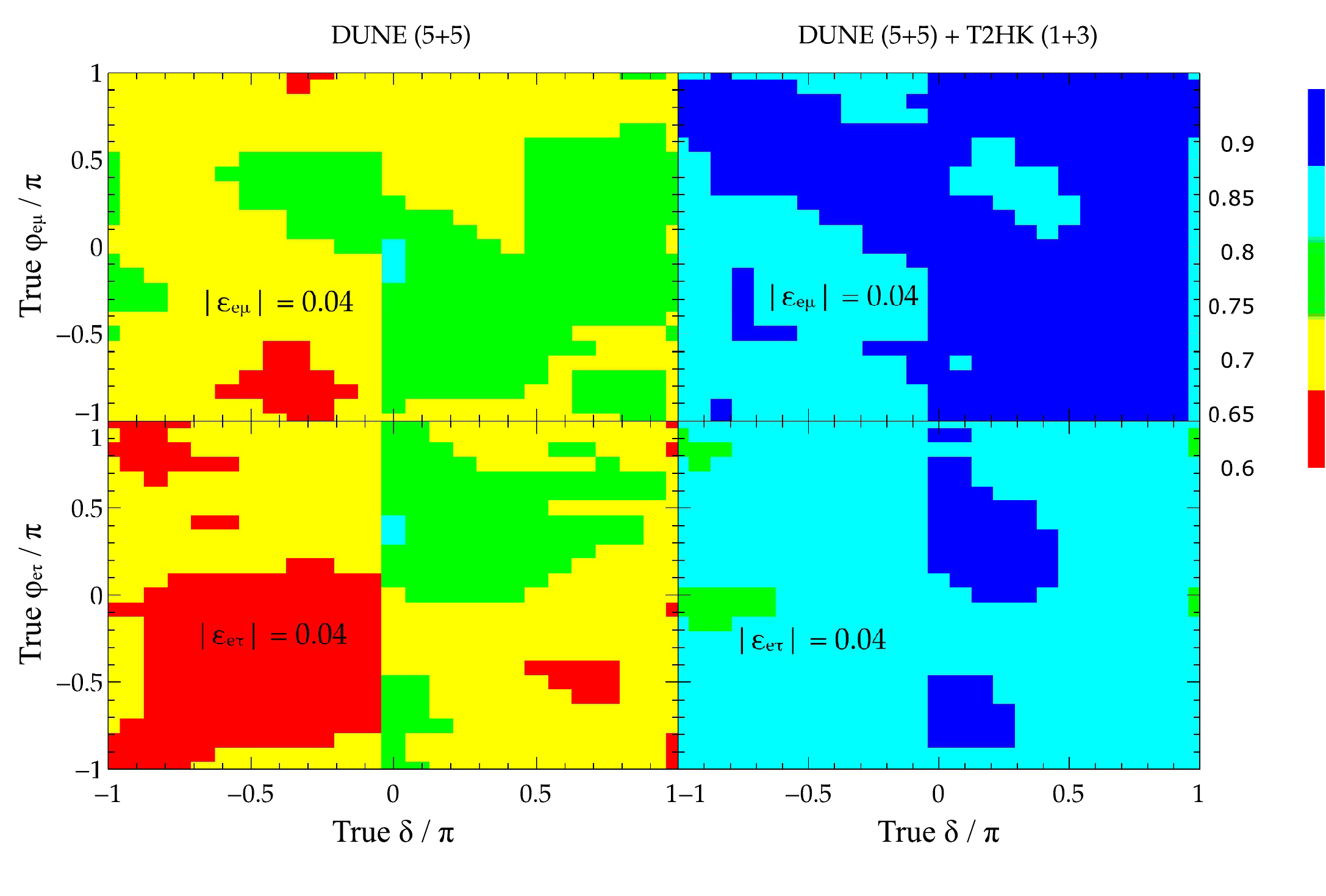}
\caption{\footnotesize{Oscillograms of generalized CP fraction in  $\varphi^{true}_{e\mu} - \delta^{true}$ plane.
 }}
\label{fig:osc}
\end{figure}
 
The region outside the $3\sigma$ contour represents those values of the pair of test CP phases which can be safely discarded above $3\sigma$ while reconstructing their values for the specific choice of the true pair of CP phases.
 Smaller enclosed regions by the contours (see Fig.~\ref{fig:fig13}, right panel) imply better measurement ability. 
Let us define a generalized CP fraction at a given confidence level~\footnote{This is different from the CP fraction that we have introduced earlier which involves only the Dirac CP phase ($\delta$).} as the ratio of the area outside the contours to the full area. This quantity allows us to have an idea of how well a pair of CP phases can be reconstructed at any given confidence level. Large CP fraction  implies better identification of the CP pair among the test values.
 
 To take into account all possible choices of the true pair of CP phases, we show in Fig.~\ref{fig:osc} oscillograms of generalized CP fraction in $\varphi^{true}_{e\mu} - \delta^{true}$ and $\varphi^{true}_{e\tau} - \delta^{true}$ plane. The colours represent values of the generalised CP fraction in the range $0.6-1$. 
 For the case of \dune, the range of generalised CP fraction is $\simeq 0.6-0.8$ while if we add \ttohk\, to \dune, the range of CP fraction becomes $\simeq 0.85-1$.  This means that \ttohk\, when combined with \dune\, allows us to measure the CP phases much better.

 %

\section{Discussion and Conclusion}
\label{sec:conclude}

With tremendous progress on both theoretical and experimental fronts in
 neutrino oscillation physics, we have fairly good idea of the neutrino masses and mixing pattern. 
 The proposed long baseline experiment, \dune\, aims to hunt for the most sought after parameter, 
 the CP violating phase $\delta$. In the era of precision, the subdominant effects due to new 
 physics such as NSI, need to be incorporated carefully. We have discussed the impact of 
 propagation NSI on the standard procedure to determine the CP violation sensitivity at long baseline experiments.  
The impact of NSI (including new CP phases) on the CP measurements at \dune\, using only the appearance channel was studied analytically as well as numerically at the level of probability and event rates in Ref.~\cite{Masud:2015xva}. In the present article, we perform a full-fledged sensitivity analysis using  appearance and disappearance channels and quantify the effects at the level of $\chi^2$ and CP fraction $f (\sigma > 3)$.
 We considered the NSI terms  individually first and then a combination of the dominant ones. 
We also compare the CP sensitivities with other ongoing  experiments - \ttok\,, \nova\, and a future generation experiment - \ttohk. 

We have found a general rule (in Sec.~\ref{results_a}) that allows us to comprehend the results very nicely.   
   There are two opposing effects at work - one is purely statistical which tends to decrease the CP sensitivity while the other one is due to the variation of the true values of the additional CP phases which tends to broaden the sensitivity bands provided the true value of the modulus of the relevant NSI parameter is large. 
    For diagonal NSI terms, only the first effect dominates while for the off-diagonal NSI terms, the additional CP phases may enhance or lower the sensitivity to CP violation thereby broadening the grey bands as well as  increasing the value of CP fraction depending on the strength of the NSI term. We would  like to stress that although the value of $\chi^{2}$ in presence of NSI may sometimes be greater than the SI case (due to effect (b)), this is not to be interpreted in a positive sense {\it i.e.}, this {does not} mean that the sensitivity to violation of the Dirac CP phase has necessarily increased. Rather it illustrates that in the presence  of new sources of CP violation (NSI moduli and phases), the sensitivity to the standard Dirac CP violation has been compromised in a very significant way.

It is shown that \dune\, is sensitive not only to CP violation effects due to the genuine SI CP phase~\cite{Bass:2013vcg} but also to additional (fake and genuine) CP violating effects arising due to  moduli and phases of the NSI parameters.
 Finally, we can infer  the following :
\begin{itemize}
\item The NSI parameters $\eem$, $\eet$ and  $\eee$ show the largest effect in the $P_{\mu e}$ channel. While the NSI parameters $\emt$, $\emm$  and $\ett$ are expected to contribute to the  $P_{\mu\mu}$ channel, we do not consider them here because the disappearance channel by itself contributes very little to the CP sensitivity owing to the absence of CP odd term in the probability. Even if we consider these parameters, we find that the $\emm$ is constrained very strongly (Eq.~\ref{tinynsi}). We have checked the impact of $\emt$ on the 
CP violation sensitivity using both appearance and disappearance channels and it is found to be negligible 
for \dune. The effect of $\ett$ is shown in Fig.~\ref{fig:ee} and is smaller in comparison to $\eee$. For these reasons, we consider $\eem$, $\eet$ and  $\eee$ as the dominant NSI parameters affecting CP sensitivity.
\item  The dependence of CP sensitivity on the  true values of $\theta_{23}$ and $\delta m^2_{31}$  
  is depicted in Fig.~\ref{fig:th23d31}. While $\theta_{23}$ changes the sensitivity both for SI and NSI, the $\delta m^2_{31}$ has minuscule effect. 
   Since $\theta_{13}$ is measured very precisely~\cite{Gonzalez-Garcia:2014bfa}, we don't expect any change due to the effect of variation of the true value of  $
\theta_{13}$.
\item We also compared the expected sensitivity of \dune\, with current long baseline experiments \ttok\,, \nova\, and  a future  experiment, \ttohk. The long baseline experiments - \ttok\, and \nova\, are not useful from the point of view of CP sensitivity both in SI and NSI when considered in isolation but adding data to \dune\, can lead to an increase in sensitivity. Inspite of the short baseline of the future \ttohk\, experiment, the high statistics leads to sensitivity that is comparable to that obtained with longer baseline experiment, \dune. It is suggested~\cite{Coloma:2015kiu} that a combination of \dune\, and \ttohk\, may be able to improve constraints on some of the NSI parameters and may also be able to resolve degeneracies.

\item The impact of NSI on the CP fraction for which sensitivity to CP violation 
is greater than $3\sigma$ at \dune\, as a function of exposure is shown 
in Fig.~\ref{fig:single_nsi_3sigma}.  
 For smaller values of the off-diagonal NSI terms, we note that the bands are always 
 below the black solid curve for SI. As we increase the value, the bands spread on either side of the black curve. 
 For diagonal NSI parameter, $\eee$ the curve due to NSI is always 
 below the black curve for the values considered here. 
  Table~\ref{tab:exposure} summarises the change in value of 
  $f (\sigma > 3)$ compared to SI case for nominal systematics
    at an optimal exposure of $350$ kt.MW.yr  in case of \dune\, for both NH and IH.

\item {{The impact of change in systematics as a function of baseline is shown in Fig.~\ref{fig:baseline_optimization} and Table~\ref{tab:system}. It is shown that the choice of optimal baseline changes when we include effects due to NSI. 
}}

\item If the source of CP violation is exactly known then the CP violation sensitivity in presence of NSI is always dominated by effect (a) as discussed in Sec.~\ref{results_f}. The results are shown in Fig.~\ref{fig:fig8} as a function of the three different phases involved.

\item The ability of  long baseline experiments to measure the value of standard or nonstandard CP phases is discussed in Sec.~\ref{results_g} and shown in Figs.~\ref{fig:fig13} and \ref{fig:osc}. Adding \ttohk\, to \dune\, helps in increasing the value of the generalised CP fraction which in turn implies that measurement of the CP phases will be far better.  

\end{itemize}


We therefore conclude that new physics effects such as propagation NSI spoil the CP sensitivity at long baseline neutrino experiments in general and specifically in the context of \dune. This spoiling can be in either direction depending upon the strength of the NSI terms involved. The problem then reduces to disentangling standard CP violation from the non-standard CP violation.

 Another  observation that can be made from this study is that 
if  $f (\sigma > 3)$ such that $f (\sigma > 3) \sim 1$ (for shorter  exposures) and 
significantly larger than its value for the given systematics, then it unequivocally 
implies  that there is new physics giving rise to it. Then the task is to find
  the new physics scenario uniquely that could give rise to such a large value. However if
   $f (\sigma > 3)$ stays far from unity for reasonable exposures 
     then we can constrain the NSI terms very well using long baseline experiments such as 
     \dune.


 If we look at different experiments, naturally the impact of NSI will be less for the relatively  shorter baselines since the amount of matter traversed will be less. So, \ttok, and \ttohk\, experience a  lesser impact than \nova\, and \dune. For shorter baselines such as $295$ km, the characteristic double peak structure of the CP sensitivity curve as a function of true $\delta$ survives and there is clear distinction between the CP conserving  and CP violating cases.
\ttohk\, due to the shorter baseline and high statistics 
does better in demarcating the CP conserving and CP violating values as the grey bands are not so broad.
 However for longer baselines such as for \nova\,  and \dune, the CP sensitivity as a function of $\delta$ flattens out for some choice of phases of the relevant NSI parameters (upper edge of the grey band). This poses a problem in differentiating the CP conserving and CP violating
  cases. In effect, this can mask/obscure the measurement of SI CP phase, 
  if the nature had  it to be CP conserving. For favourable values  of the CP phase (near $\pm \pi/2$), 
  \dune\, will be competitive with \ttohk. But, if CP phases were in the 
  unfavourable region  (near $\sim 0, \pi$) , then the future \ttohk\, experiment  will be very promising.

\section*{Acknowledgements} 
It is a pleasure to thank Raj Gandhi for useful discussions and critical comments on the manuscript.   
 We acknowledge the use of HRI cluster facility to carry out computations in this work.
PM acknowledges support from University Grants Commission under the second phase of University with Potential of Excellence at JNU and partial support from the European Union grant FP7 ITN INVISIBLES (Marie Curie Actions, PITN-GA-2011-289442). We thank the organisers of WHEPP at IIT-Kanpur for the warm hospitality where substantial progress in the work was carried out. We are grateful to the anonymous referee for constructive suggestions and inputs.

\bibliographystyle{apsrev}
\bibliography{referencesnsi}

\begin{thebibliography}{10}
\expandafter\ifx\csname bibnamefont\endcsname\relax
  \def\bibnamefont#1{#1}\fi
\expandafter\ifx\csname bibfnamefont\endcsname\relax
  \def\bibfnamefont#1{#1}\fi
\expandafter\ifx\csname url\endcsname\relax
  \def\url#1{\texttt{#1}}\fi
\expandafter\ifx\csname urlprefix\endcsname\relax\def\urlprefix{URL }\fi
\providecommand{\bibinfo}[2]{#2}
\providecommand{\eprint}[2][]{\url{#2}}

\bibitem{Wolfenstein:1964ks}
\bibinfo{author}{\bibfnamefont{L.}~\bibnamefont{Wolfenstein}},
  \bibinfo{journal}{Phys. Rev. Lett.} \textbf{\bibinfo{volume}{13}},
  \bibinfo{pages}{562} (\bibinfo{year}{1964}).

\bibitem{Branco:2011zb}
\bibinfo{author}{\bibfnamefont{G.~C.} \bibnamefont{Branco}},
  \bibinfo{author}{\bibfnamefont{R.~G.} \bibnamefont{Felipe}},
  \bibnamefont{and} \bibinfo{author}{\bibfnamefont{F.~R.}
  \bibnamefont{Joaquim}}, \bibinfo{journal}{Rev. Mod. Phys.}
  \textbf{\bibinfo{volume}{84}}, \bibinfo{pages}{515} (\bibinfo{year}{2012}),
  \eprint{1111.5332}.

\bibitem{km}
\bibinfo{author}{\bibfnamefont{M.}~\bibnamefont{Kobayashi}} \bibnamefont{and}
  \bibinfo{author}{\bibfnamefont{T.}~\bibnamefont{Maskawa}},
  \bibinfo{journal}{Progress of Theoretical Physics}
  \textbf{\bibinfo{volume}{49}}(\bibinfo{number}{2}), \bibinfo{pages}{652}
  (\bibinfo{year}{1973}).

\bibitem{Barger:1980jm}
\bibinfo{author}{\bibfnamefont{V.~D.} \bibnamefont{Barger}},
  \bibinfo{author}{\bibfnamefont{K.}~\bibnamefont{Whisnant}}, \bibnamefont{and}
  \bibinfo{author}{\bibfnamefont{R.~J.~N.} \bibnamefont{Phillips}},
  \bibinfo{journal}{Phys. Rev. Lett.} \textbf{\bibinfo{volume}{45}},
  \bibinfo{pages}{2084} (\bibinfo{year}{1980}).

\bibitem{Nunokawa:2007qh}
\bibinfo{author}{\bibfnamefont{H.}~\bibnamefont{Nunokawa}},
  \bibinfo{author}{\bibfnamefont{S.~J.} \bibnamefont{Parke}}, \bibnamefont{and}
  \bibinfo{author}{\bibfnamefont{J.~W.~F.} \bibnamefont{Valle}},
  \bibinfo{journal}{Prog. Part. Nucl. Phys.} \textbf{\bibinfo{volume}{60}},
  \bibinfo{pages}{338} (\bibinfo{year}{2008}), \eprint{0710.0554}.

\bibitem{Farzan:2006vj}
\bibinfo{author}{\bibfnamefont{Y.}~\bibnamefont{Farzan}} \bibnamefont{and}
  \bibinfo{author}{\bibfnamefont{A.~{\relax Yu}.} \bibnamefont{Smirnov}},
  \bibinfo{journal}{JHEP} \textbf{\bibinfo{volume}{01}}, \bibinfo{pages}{059}
  (\bibinfo{year}{2007}), \eprint{hep-ph/0610337}.

\bibitem{Arafune:1997hd}
\bibinfo{author}{\bibfnamefont{J.}~\bibnamefont{Arafune}},
  \bibinfo{author}{\bibfnamefont{M.}~\bibnamefont{Koike}}, \bibnamefont{and}
  \bibinfo{author}{\bibfnamefont{J.}~\bibnamefont{Sato}},
  \bibinfo{journal}{Phys. Rev.} \textbf{\bibinfo{volume}{D56}},
  \bibinfo{pages}{3093} (\bibinfo{year}{1997}), \bibinfo{note}{[Erratum: Phys.
  Rev.D60,119905(1999)]}, \eprint{hep-ph/9703351}.

\bibitem{Ohlsson:2013ip}
\bibinfo{author}{\bibfnamefont{T.}~\bibnamefont{Ohlsson}},
  \bibinfo{author}{\bibfnamefont{H.}~\bibnamefont{Zhang}}, \bibnamefont{and}
  \bibinfo{author}{\bibfnamefont{S.}~\bibnamefont{Zhou}},
  \bibinfo{journal}{Phys. Rev.}
  \textbf{\bibinfo{volume}{D87}}(\bibinfo{number}{5}), \bibinfo{pages}{053006}
  (\bibinfo{year}{2013}), \eprint{1301.4333}.

\bibitem{Marciano:2006uc}
\bibinfo{author}{\bibfnamefont{W.}~\bibnamefont{Marciano}} \bibnamefont{and}
  \bibinfo{author}{\bibfnamefont{Z.}~\bibnamefont{Parsa}},
  \bibinfo{journal}{Nucl. Phys. Proc. Suppl.} \textbf{\bibinfo{volume}{221}},
  \bibinfo{pages}{166} (\bibinfo{year}{2011}), \eprint{hep-ph/0610258}.

\bibitem{Bass:2013vcg}
\bibinfo{author}{\bibfnamefont{M.}~\bibnamefont{Bass}} \emph{et~al.}
  (\bibinfo{collaboration}{LBNE Collaboration}), \bibinfo{journal}{Phys. Rev.}
  \textbf{\bibinfo{volume}{D91}}, \bibinfo{pages}{052015}
  (\bibinfo{year}{2015}), \eprint{1311.0212}.

\bibitem{Acciarri:2015uup}
\bibinfo{author}{\bibfnamefont{R.}~\bibnamefont{Acciarri}} \emph{et~al.}
  (\bibinfo{collaboration}{DUNE})  (\bibinfo{year}{2015}), \eprint{1512.06148}.

\bibitem{Acciarri:2016ooe}
\bibinfo{author}{\bibfnamefont{R.}~\bibnamefont{Acciarri}} \emph{et~al.}
  (\bibinfo{collaboration}{DUNE})  (\bibinfo{year}{2016}), \eprint{1601.02984}.

\bibitem{Acciarri:2016crz}
\bibinfo{author}{\bibfnamefont{R.}~\bibnamefont{Acciarri}} \emph{et~al.}
  (\bibinfo{collaboration}{DUNE})  (\bibinfo{year}{2016}), \eprint{1601.05471}.

\bibitem{Qian:2015waa}
\bibinfo{author}{\bibfnamefont{X.}~\bibnamefont{Qian}} \bibnamefont{and}
  \bibinfo{author}{\bibfnamefont{P.}~\bibnamefont{Vogel}},
  \bibinfo{journal}{Prog. Part. Nucl. Phys.} \textbf{\bibinfo{volume}{83}},
  \bibinfo{pages}{1} (\bibinfo{year}{2015}), \eprint{1505.01891}.

\bibitem{Barger:2013rha}
\bibinfo{author}{\bibfnamefont{V.}~\bibnamefont{Barger}},
  \bibinfo{author}{\bibfnamefont{A.}~\bibnamefont{Bhattacharya}},
  \bibinfo{author}{\bibfnamefont{A.}~\bibnamefont{Chatterjee}},
  \bibinfo{author}{\bibfnamefont{R.}~\bibnamefont{Gandhi}},
  \bibinfo{author}{\bibfnamefont{D.}~\bibnamefont{Marfatia}}, \bibnamefont{and}
  \bibinfo{author}{\bibfnamefont{M.}~\bibnamefont{Masud}},
  \bibinfo{journal}{Phys. Rev.}
  \textbf{\bibinfo{volume}{D89}}(\bibinfo{number}{1}), \bibinfo{pages}{011302}
  (\bibinfo{year}{2014}), \eprint{1307.2519}.

\bibitem{Barger:2014dfa}
\bibinfo{author}{\bibfnamefont{V.}~\bibnamefont{Barger}},
  \bibinfo{author}{\bibfnamefont{A.}~\bibnamefont{Bhattacharya}},
  \bibinfo{author}{\bibfnamefont{A.}~\bibnamefont{Chatterjee}},
  \bibinfo{author}{\bibfnamefont{R.}~\bibnamefont{Gandhi}},
  \bibinfo{author}{\bibfnamefont{D.}~\bibnamefont{Marfatia}}, \bibnamefont{and}
  \bibinfo{author}{\bibfnamefont{M.}~\bibnamefont{Masud}},
  \bibinfo{journal}{Int. J. Mod. Phys.}
  \textbf{\bibinfo{volume}{A31}}(\bibinfo{number}{07}),
  \bibinfo{pages}{1650020} (\bibinfo{year}{2016}), \eprint{1405.1054}.

\bibitem{Ghosh:2014rna}
\bibinfo{author}{\bibfnamefont{M.}~\bibnamefont{Ghosh}},
  \bibinfo{author}{\bibfnamefont{S.}~\bibnamefont{Goswami}}, \bibnamefont{and}
  \bibinfo{author}{\bibfnamefont{S.~K.} \bibnamefont{Raut}},
  \bibinfo{journal}{Eur. Phys. J.}
  \textbf{\bibinfo{volume}{C76}}(\bibinfo{number}{3}), \bibinfo{pages}{114}
  (\bibinfo{year}{2016}), \eprint{1412.1744}.

\bibitem{Abe:2015zbg}
\bibinfo{author}{\bibfnamefont{K.}~\bibnamefont{Abe}} \emph{et~al.}
  (\bibinfo{collaboration}{Hyper-Kamiokande Proto-Collaboration}),
  \bibinfo{journal}{PTEP} \textbf{\bibinfo{volume}{2015}},
  \bibinfo{pages}{053C02} (\bibinfo{year}{2015}), \eprint{1502.05199}.

\bibitem{Blennow:2015cmn}
\bibinfo{author}{\bibfnamefont{M.}~\bibnamefont{Blennow}},
  \bibinfo{author}{\bibfnamefont{P.}~\bibnamefont{Coloma}}, \bibnamefont{and}
  \bibinfo{author}{\bibfnamefont{E.}~\bibnamefont{Fernández-Martinez}}
  (\bibinfo{year}{2015}), \eprint{1511.02859}.

\bibitem{Farzan:2015doa}
\bibinfo{author}{\bibfnamefont{Y.}~\bibnamefont{Farzan}},
  \bibinfo{journal}{Phys. Lett.} \textbf{\bibinfo{volume}{B748}},
  \bibinfo{pages}{311} (\bibinfo{year}{2015}), \eprint{1505.06906}.

\bibitem{Farzan:2015hkd}
\bibinfo{author}{\bibfnamefont{Y.}~\bibnamefont{Farzan}} \bibnamefont{and}
  \bibinfo{author}{\bibfnamefont{I.~M.} \bibnamefont{Shoemaker}}
  (\bibinfo{year}{2015}), \eprint{1512.09147}.

\bibitem{GonzalezGarcia:2001mp}
\bibinfo{author}{\bibfnamefont{M.~C.} \bibnamefont{Gonzalez-Garcia}},
  \bibinfo{author}{\bibfnamefont{Y.}~\bibnamefont{Grossman}},
  \bibinfo{author}{\bibfnamefont{A.}~\bibnamefont{Gusso}}, \bibnamefont{and}
  \bibinfo{author}{\bibfnamefont{Y.}~\bibnamefont{Nir}},
  \bibinfo{journal}{Phys. Rev.} \textbf{\bibinfo{volume}{D64}},
  \bibinfo{pages}{096006} (\bibinfo{year}{2001}), \eprint{hep-ph/0105159}.

\bibitem{Winter:2008eg}
\bibinfo{author}{\bibfnamefont{W.}~\bibnamefont{Winter}},
  \bibinfo{journal}{Phys. Lett.} \textbf{\bibinfo{volume}{B671}},
  \bibinfo{pages}{77} (\bibinfo{year}{2009}), \eprint{0808.3583}.

\bibitem{Chatterjee:2014gxa}
\bibinfo{author}{\bibfnamefont{A.}~\bibnamefont{Chatterjee}},
  \bibinfo{author}{\bibfnamefont{P.}~\bibnamefont{Mehta}},
  \bibinfo{author}{\bibfnamefont{D.}~\bibnamefont{Choudhury}},
  \bibnamefont{and} \bibinfo{author}{\bibfnamefont{R.}~\bibnamefont{Gandhi}}
  (\bibinfo{year}{2014}), \eprint{1409.8472}.

\bibitem{Forero:2014bxa}
\bibinfo{author}{\bibfnamefont{D.~V.} \bibnamefont{Forero}},
  \bibinfo{author}{\bibfnamefont{M.}~\bibnamefont{Tortola}}, \bibnamefont{and}
  \bibinfo{author}{\bibfnamefont{J.~W.~F.} \bibnamefont{Valle}},
  \bibinfo{journal}{Phys. Rev.}
  \textbf{\bibinfo{volume}{D90}}(\bibinfo{number}{9}), \bibinfo{pages}{093006}
  (\bibinfo{year}{2014}), \eprint{1405.7540}.

\bibitem{Abe:2015awa}
\bibinfo{author}{\bibfnamefont{K.}~\bibnamefont{Abe}} \emph{et~al.}
  (\bibinfo{collaboration}{T2K}), \bibinfo{journal}{Phys. Rev.}
  \textbf{\bibinfo{volume}{D91}}(\bibinfo{number}{7}), \bibinfo{pages}{072010}
  (\bibinfo{year}{2015}), \eprint{1502.01550}.

\bibitem{Adamson:2016tbq}
\bibinfo{author}{\bibfnamefont{P.}~\bibnamefont{Adamson}} \emph{et~al.}
  (\bibinfo{collaboration}{NOvA})  (\bibinfo{year}{2016}), \eprint{1601.05022}.

\bibitem{sktalk}
\bibinfo{author}{\bibfnamefont{C.}~\bibnamefont{Kachulis}},
  \emph{\bibinfo{title}{Sk atmospheric neutrino results}},
  \bibinfo{note}{https://indico.cern.ch/event/361123/session/2/
  contribution/348/attachments/1136004/1625868/SK\_atmospheric\_kachulis\_dpf2015.pdf}.

\bibitem{Forero:2016cmb}
\bibinfo{author}{\bibfnamefont{D.~V.} \bibnamefont{Forero}} \bibnamefont{and}
  \bibinfo{author}{\bibfnamefont{P.}~\bibnamefont{Huber}}
  (\bibinfo{year}{2016}), \eprint{1601.03736}.

\bibitem{Ohlsson:2012kf}
\bibinfo{author}{\bibfnamefont{T.}~\bibnamefont{Ohlsson}},
  \bibinfo{journal}{Rept. Prog. Phys.} \textbf{\bibinfo{volume}{76}},
  \bibinfo{pages}{044201} (\bibinfo{year}{2013}), \eprint{1209.2710}.

\bibitem{Miranda:2015dra}
\bibinfo{author}{\bibfnamefont{O.~G.} \bibnamefont{Miranda}} \bibnamefont{and}
  \bibinfo{author}{\bibfnamefont{H.}~\bibnamefont{Nunokawa}},
  \bibinfo{journal}{New J. Phys.}
  \textbf{\bibinfo{volume}{17}}(\bibinfo{number}{9}), \bibinfo{pages}{095002}
  (\bibinfo{year}{2015}), \eprint{1505.06254}.

\bibitem{Datta2004356}
\bibinfo{author}{\bibfnamefont{A.}~\bibnamefont{Datta}},
  \bibinfo{author}{\bibfnamefont{R.}~\bibnamefont{Gandhi}},
  \bibinfo{author}{\bibfnamefont{P.}~\bibnamefont{Mehta}}, \bibnamefont{and}
  \bibinfo{author}{\bibfnamefont{S.~U.} \bibnamefont{Sankar}},
  \bibinfo{journal}{Physics Letters B}
  \textbf{\bibinfo{volume}{597}}(\bibinfo{number}{3–4}), \bibinfo{pages}{356
  } (\bibinfo{year}{2004}).

\bibitem{Chatterjee:2014oda}
\bibinfo{author}{\bibfnamefont{A.}~\bibnamefont{Chatterjee}},
  \bibinfo{author}{\bibfnamefont{R.}~\bibnamefont{Gandhi}}, \bibnamefont{and}
  \bibinfo{author}{\bibfnamefont{J.}~\bibnamefont{Singh}},
  \bibinfo{journal}{JHEP} \textbf{\bibinfo{volume}{1406}}, \bibinfo{pages}{045}
  (\bibinfo{year}{2014}), \eprint{1402.6265}.

\bibitem{Gandhi:2015xza}
\bibinfo{author}{\bibfnamefont{R.}~\bibnamefont{Gandhi}},
  \bibinfo{author}{\bibfnamefont{B.}~\bibnamefont{Kayser}},
  \bibinfo{author}{\bibfnamefont{M.}~\bibnamefont{Masud}}, \bibnamefont{and}
  \bibinfo{author}{\bibfnamefont{S.}~\bibnamefont{Prakash}},
  \bibinfo{journal}{JHEP} \textbf{\bibinfo{volume}{11}}, \bibinfo{pages}{039}
  (\bibinfo{year}{2015}), \eprint{1508.06275}.

\bibitem{Berryman:2015nua}
\bibinfo{author}{\bibfnamefont{J.~M.} \bibnamefont{Berryman}},
  \bibinfo{author}{\bibfnamefont{A.}~\bibnamefont{de~Gouva}},
  \bibinfo{author}{\bibfnamefont{K.~J.} \bibnamefont{Kelly}}, \bibnamefont{and}
  \bibinfo{author}{\bibfnamefont{A.}~\bibnamefont{Kobach}},
  \bibinfo{journal}{Phys. Rev.}
  \textbf{\bibinfo{volume}{D92}}(\bibinfo{number}{7}), \bibinfo{pages}{073012}
  (\bibinfo{year}{2015}), \eprint{1507.03986}.

\bibitem{Masud:2015xva}
\bibinfo{author}{\bibfnamefont{M.}~\bibnamefont{Masud}},
  \bibinfo{author}{\bibfnamefont{A.}~\bibnamefont{Chatterjee}},
  \bibnamefont{and} \bibinfo{author}{\bibfnamefont{P.}~\bibnamefont{Mehta}}
  (\bibinfo{year}{2015}), \eprint{1510.08261}.

\bibitem{Friedland:2012tq}
\bibinfo{author}{\bibfnamefont{A.}~\bibnamefont{Friedland}} \bibnamefont{and}
  \bibinfo{author}{\bibfnamefont{I.~M.} \bibnamefont{Shoemaker}}
  (\bibinfo{year}{2012}), \eprint{1207.6642}.

\bibitem{Coloma:2015kiu}
\bibinfo{author}{\bibfnamefont{P.}~\bibnamefont{Coloma}},
  \bibinfo{journal}{JHEP} \textbf{\bibinfo{volume}{03}}, \bibinfo{pages}{016}
  (\bibinfo{year}{2016}), \eprint{1511.06357}.

\bibitem{deGouvea:2015ndi}
\bibinfo{author}{\bibfnamefont{A.}~\bibnamefont{de~Gouvea}} \bibnamefont{and}
  \bibinfo{author}{\bibfnamefont{K.~J.} \bibnamefont{Kelly}}
  (\bibinfo{year}{2015}), \eprint{1511.05562}.

\bibitem{Liao:2016hsa}
\bibinfo{author}{\bibfnamefont{J.}~\bibnamefont{Liao}},
  \bibinfo{author}{\bibfnamefont{D.}~\bibnamefont{Marfatia}}, \bibnamefont{and}
  \bibinfo{author}{\bibfnamefont{K.}~\bibnamefont{Whisnant}}
  (\bibinfo{year}{2016}), \eprint{1601.00927}.

\bibitem{Huitu:2016bmb}
\bibinfo{author}{\bibfnamefont{K.}~\bibnamefont{Huitu}},
  \bibinfo{author}{\bibfnamefont{T.~J.} \bibnamefont{KŠrkkŠinen}},
  \bibinfo{author}{\bibfnamefont{J.}~\bibnamefont{Maalampi}}, \bibnamefont{and}
  \bibinfo{author}{\bibfnamefont{S.}~\bibnamefont{Vihonen}}
  (\bibinfo{year}{2016}), \eprint{1601.07730}.

\bibitem{Bakhti:2016prn}
\bibinfo{author}{\bibfnamefont{P.}~\bibnamefont{Bakhti}} \bibnamefont{and}
  \bibinfo{author}{\bibfnamefont{Y.}~\bibnamefont{Farzan}}
  (\bibinfo{year}{2016}), \eprint{1602.07099}.

\bibitem{Beringer:1900zz}
\bibinfo{author}{\bibfnamefont{J.}~\bibnamefont{Beringer}} \emph{et~al.}
  (\bibinfo{collaboration}{Particle Data Group}), \bibinfo{journal}{Phys. Rev.}
  \textbf{\bibinfo{volume}{D86}}, \bibinfo{pages}{010001}
  (\bibinfo{year}{2012}).

\bibitem{Mehta:2009ea}
\bibinfo{author}{\bibfnamefont{P.}~\bibnamefont{Mehta}},
  \bibinfo{journal}{Phys. Rev.} \textbf{\bibinfo{volume}{D79}},
  \bibinfo{pages}{096013} (\bibinfo{year}{2009}), \eprint{0901.0790}.

\bibitem{Mehta:2009xm}
\bibinfo{author}{\bibfnamefont{P.}~\bibnamefont{Mehta}}
  (\bibinfo{year}{2009}), \eprint{0907.0562}.

\bibitem{Kikuchi:2008vq}
\bibinfo{author}{\bibfnamefont{T.}~\bibnamefont{Kikuchi}},
  \bibinfo{author}{\bibfnamefont{H.}~\bibnamefont{Minakata}}, \bibnamefont{and}
  \bibinfo{author}{\bibfnamefont{S.}~\bibnamefont{Uchinami}},
  \bibinfo{journal}{JHEP} \textbf{\bibinfo{volume}{0903}}, \bibinfo{pages}{114}
  (\bibinfo{year}{2009}), \eprint{0809.3312}.

\bibitem{Huber:2004ka}
\bibinfo{author}{\bibfnamefont{P.}~\bibnamefont{Huber}},
  \bibinfo{author}{\bibfnamefont{M.}~\bibnamefont{Lindner}}, \bibnamefont{and}
  \bibinfo{author}{\bibfnamefont{W.}~\bibnamefont{Winter}},
  \bibinfo{journal}{Comput. Phys. Commun.} \textbf{\bibinfo{volume}{167}},
  \bibinfo{pages}{195} (\bibinfo{year}{2005}), \eprint{hep-ph/0407333}.

\bibitem{Kopp:2006wp}
\bibinfo{author}{\bibfnamefont{J.}~\bibnamefont{Kopp}}, \bibinfo{journal}{Int.
  J. Mod. Phys.} \textbf{\bibinfo{volume}{C19}}, \bibinfo{pages}{523}
  (\bibinfo{year}{2008}), \eprint{physics/0610206}.

\bibitem{Huber:2007ji}
\bibinfo{author}{\bibfnamefont{P.}~\bibnamefont{Huber}},
  \bibinfo{author}{\bibfnamefont{J.}~\bibnamefont{Kopp}},
  \bibinfo{author}{\bibfnamefont{M.}~\bibnamefont{Lindner}},
  \bibinfo{author}{\bibfnamefont{M.}~\bibnamefont{Rolinec}}, \bibnamefont{and}
  \bibinfo{author}{\bibfnamefont{W.}~\bibnamefont{Winter}},
  \bibinfo{journal}{Comput. Phys. Commun.} \textbf{\bibinfo{volume}{177}},
  \bibinfo{pages}{432} (\bibinfo{year}{2007}), \eprint{hep-ph/0701187}.

\bibitem{Kopp:2007ne}
\bibinfo{author}{\bibfnamefont{J.}~\bibnamefont{Kopp}},
  \bibinfo{author}{\bibfnamefont{M.}~\bibnamefont{Lindner}},
  \bibinfo{author}{\bibfnamefont{T.}~\bibnamefont{Ota}}, \bibnamefont{and}
  \bibinfo{author}{\bibfnamefont{J.}~\bibnamefont{Sato}},
  \bibinfo{journal}{Phys. Rev.} \textbf{\bibinfo{volume}{D77}},
  \bibinfo{pages}{013007} (\bibinfo{year}{2008}), \eprint{0708.0152}.

\bibitem{Dziewonski:1981xy}
\bibinfo{author}{\bibfnamefont{A.~M.} \bibnamefont{Dziewonski}}
  \bibnamefont{and} \bibinfo{author}{\bibfnamefont{D.~L.}
  \bibnamefont{Anderson}}, \bibinfo{journal}{Phys. Earth Planet. Interiors}
  \textbf{\bibinfo{volume}{25}}, \bibinfo{pages}{297} (\bibinfo{year}{1981}).

\bibitem{Gandhi:2004md}
\bibinfo{author}{\bibfnamefont{R.}~\bibnamefont{Gandhi}},
  \bibinfo{author}{\bibfnamefont{P.}~\bibnamefont{Ghoshal}},
  \bibinfo{author}{\bibfnamefont{S.}~\bibnamefont{Goswami}},
  \bibinfo{author}{\bibfnamefont{P.}~\bibnamefont{Mehta}}, \bibnamefont{and}
  \bibinfo{author}{\bibfnamefont{S.~U.} \bibnamefont{Sankar}},
  \bibinfo{journal}{Phys.Rev.Lett.} \textbf{\bibinfo{volume}{94}},
  \bibinfo{pages}{051801} (\bibinfo{year}{2005}), \eprint{hep-ph/0408361}.

\bibitem{Gandhi:2004bj}
\bibinfo{author}{\bibfnamefont{R.}~\bibnamefont{Gandhi}},
  \bibinfo{author}{\bibfnamefont{P.}~\bibnamefont{Ghoshal}},
  \bibinfo{author}{\bibfnamefont{S.}~\bibnamefont{Goswami}},
  \bibinfo{author}{\bibfnamefont{P.}~\bibnamefont{Mehta}}, \bibnamefont{and}
  \bibinfo{author}{\bibfnamefont{S.~U.} \bibnamefont{Sankar}},
  \bibinfo{journal}{Phys.Rev.} \textbf{\bibinfo{volume}{D73}},
  \bibinfo{pages}{053001} (\bibinfo{year}{2006}), \eprint{hep-ph/0411252}.

\bibitem{GonzalezGarcia:2012sz}
\bibinfo{author}{\bibfnamefont{M.}~\bibnamefont{Gonzalez-Garcia}},
  \bibinfo{author}{\bibfnamefont{M.}~\bibnamefont{Maltoni}},
  \bibinfo{author}{\bibfnamefont{J.}~\bibnamefont{Salvado}}, \bibnamefont{and}
  \bibinfo{author}{\bibfnamefont{T.}~\bibnamefont{Schwetz}},
  \bibinfo{journal}{JHEP} \textbf{\bibinfo{volume}{1212}}, \bibinfo{pages}{123}
  (\bibinfo{year}{2012}), \eprint{1209.3023}.

\bibitem{Capozzi:2013csa}
\bibinfo{author}{\bibfnamefont{F.}~\bibnamefont{Capozzi}},
  \bibinfo{author}{\bibfnamefont{G.}~\bibnamefont{Fogli}},
  \bibinfo{author}{\bibfnamefont{E.}~\bibnamefont{Lisi}},
  \bibinfo{author}{\bibfnamefont{A.}~\bibnamefont{Marrone}},
  \bibinfo{author}{\bibfnamefont{D.}~\bibnamefont{Montanino}}, \emph{et~al.},
  \bibinfo{journal}{Phys.Rev.} \textbf{\bibinfo{volume}{D89}},
  \bibinfo{pages}{093018} (\bibinfo{year}{2014}), \eprint{1312.2878}.

\bibitem{Kimura:2002hb}
\bibinfo{author}{\bibfnamefont{K.}~\bibnamefont{Kimura}},
  \bibinfo{author}{\bibfnamefont{A.}~\bibnamefont{Takamura}}, \bibnamefont{and}
  \bibinfo{author}{\bibfnamefont{H.}~\bibnamefont{Yokomakura}},
  \bibinfo{journal}{Phys. Lett.} \textbf{\bibinfo{volume}{B537}},
  \bibinfo{pages}{86} (\bibinfo{year}{2002}), \eprint{hep-ph/0203099}.

\bibitem{Kimura:2002wd}
\bibinfo{author}{\bibfnamefont{K.}~\bibnamefont{Kimura}},
  \bibinfo{author}{\bibfnamefont{A.}~\bibnamefont{Takamura}}, \bibnamefont{and}
  \bibinfo{author}{\bibfnamefont{H.}~\bibnamefont{Yokomakura}},
  \bibinfo{journal}{Phys. Rev.} \textbf{\bibinfo{volume}{D66}},
  \bibinfo{pages}{073005} (\bibinfo{year}{2002}), \eprint{hep-ph/0205295}.

\bibitem{Yasuda:2007jp}
\bibinfo{author}{\bibfnamefont{O.}~\bibnamefont{Yasuda}}
  (\bibinfo{year}{2007}), \eprint{0704.1531}.

\bibitem{Meloni:2009ia}
\bibinfo{author}{\bibfnamefont{D.}~\bibnamefont{Meloni}},
  \bibinfo{author}{\bibfnamefont{T.}~\bibnamefont{Ohlsson}}, \bibnamefont{and}
  \bibinfo{author}{\bibfnamefont{H.}~\bibnamefont{Zhang}},
  \bibinfo{journal}{JHEP} \textbf{\bibinfo{volume}{04}}, \bibinfo{pages}{033}
  (\bibinfo{year}{2009}), \eprint{0901.1784}.

\bibitem{Barger:2001yr}
\bibinfo{author}{\bibfnamefont{V.}~\bibnamefont{Barger}},
  \bibinfo{author}{\bibfnamefont{D.}~\bibnamefont{Marfatia}}, \bibnamefont{and}
  \bibinfo{author}{\bibfnamefont{K.}~\bibnamefont{Whisnant}},
  \bibinfo{journal}{Phys. Rev.} \textbf{\bibinfo{volume}{D65}},
  \bibinfo{pages}{073023} (\bibinfo{year}{2002}), \eprint{hep-ph/0112119}.

\bibitem{Gonzalez-Garcia:2014bfa}
\bibinfo{author}{\bibfnamefont{M.~C.} \bibnamefont{Gonzalez-Garcia}},
  \bibinfo{author}{\bibfnamefont{M.}~\bibnamefont{Maltoni}}, \bibnamefont{and}
  \bibinfo{author}{\bibfnamefont{T.}~\bibnamefont{Schwetz}},
  \bibinfo{journal}{JHEP} \textbf{\bibinfo{volume}{11}}, \bibinfo{pages}{052}
  (\bibinfo{year}{2014}), \eprint{1409.5439}.

\bibitem{Qian:2012zn}
\bibinfo{author}{\bibfnamefont{X.}~\bibnamefont{Qian}},
  \bibinfo{author}{\bibfnamefont{A.}~\bibnamefont{Tan}},
  \bibinfo{author}{\bibfnamefont{W.}~\bibnamefont{Wang}},
  \bibinfo{author}{\bibfnamefont{J.~J.} \bibnamefont{Ling}},
  \bibinfo{author}{\bibfnamefont{R.~D.} \bibnamefont{McKeown}},
  \bibnamefont{and} \bibinfo{author}{\bibfnamefont{C.}~\bibnamefont{Zhang}},
  \bibinfo{journal}{Phys. Rev.} \textbf{\bibinfo{volume}{D86}},
  \bibinfo{pages}{113011} (\bibinfo{year}{2012}), \eprint{1210.3651}.

\end{thebibliography}

\end{document}